\documentclass[
  aps,
  prl,
  reprint,
  10pt,
  superscriptaddress,
  nofootinbib,
  amsfonts,
  amsmath,
  amssymb
]{revtex4-2}

\usepackage{bm}
\usepackage{bbold}
\usepackage{braket}

\usepackage{booktabs}

\usepackage[version=4]{mhchem}
\usepackage{microtype}
\usepackage{upgreek}
\usepackage{graphicx}
\usepackage{hyperref}
\usepackage{xcolor}
\hypersetup{
  colorlinks,
  linkcolor={blue!80!black},
  citecolor={blue!80!black},
  urlcolor={blue!80!black},
  pdftitle={Emergent Gauge Fields and Linear Dichroism in
  Spin-Textured Altermagnets},
  pdfauthor={Andrea Maiani}
}

\begin{document}

\title{Emergent Gauge Fields and Linear Dichroism in Spin-Textured Altermagnets}

\author{Andrea Maiani}
\affiliation{Nordita, KTH Royal Institute of Technology and Stockholm University,
Hannes Alfvéns väg 12, SE-10691 Stockholm, Sweden}

\begin{abstract}
  Spin textures are ubiquitous in antiferromagnets,
  yet their consequences for altermagnets remain largely unexplored.
  We show that smooth spatial variations of the N\'eel order
  act on itinerant electrons as emergent gauge fields
  that couple to the altermagnetic sublattice-odd channel.
  For a coplanar spin helix,
  this channel generates two responses that vanish when it is removed:
  a momentum-resolved sublattice-odd spectral structure and a net interband linear dichroism
  whose principal axes are controlled by the helix wave vector,
  so the optical response can be steered by the magnetic texture.
  The dichroism exhibits two distinct regimes separated by a crossover:
  at low frequencies the absorption axis is locked to the crystal axes,
  while at high frequencies it tracks the helix.
  Anisotropic dc transport provides a complementary quantitative signature,
  and all signatures are robust to weak intrinsic spin-orbit coupling.
  Our results identify polarization-resolved optics
  and sublattice-sensitive spectroscopy
  as direct probes of textured altermagnetic states
  and suggest a route
  to direction-selective electronic and optical functionality in altermagnets.
\end{abstract}

\date{20 July 2026}
\maketitle

Altermagnets are antiferromagnetically ordered materials with vanishing net magnetization
where opposite-spin sublattices are related by a crystal symmetry other
than a simple real-space translation or inversion~\cite{Hayami_2019_Momentum,
  Hayami_2020_Spontaneous, Naka_2019_Spin, Naka_2020_Anomalous, Smejkal_2020_Crystal,
  Mazin_2022_Editorial, Smejkal_2022_Emerging, Olsen_2024_Antiferromagnetism,
Autieri_2024_New}.
This multipolar order breaks time-reversal and crystal-rotation
symmetries separately while preserving their combinations, yielding spin-split bands at
finite momentum despite zero net magnetization.
Experimental signatures have been
reported in materials including \ce{RuO2}, \ce{\alpha-MnTe}, and \ce{Mn5Si3}, among
others~\cite{Ahn_2019_Antiferromagnetism, Smejkal_2022_Beyond, Lee_2024_Broken,
  Osumi_2024_Observation, Reichlova_2024_Observation, Krempasky_2024_Altermagnetic,
Reimers_2024_Direct, Fernandes_2024_Topological}.
In particular, layered $d$-wave altermagnets in the \ce{V2Se2O} family,
including \ce{KV2Se2O} and \ce{Rb_{1-\delta}V2Te2O},
provide quasi-two-dimensional platforms with spin-resolved evidence
of room-temperature spin-momentum locking~\cite{Jiang_2025_Metallic, Zhang_2025_Crystal}.

Domain formation and nonuniform N\'eel configurations are common in antiferromagnets
due to the absence of stray fields~\cite{Gomonay_2018_Antiferromagnetic}.
Consistent with this expectation, nanoscale probes already indicate nonuniform N\'eel textures in
candidate altermagnets~\cite{Amin_2024_Nanoscale, Yamamoto_2025_Altermagnetic,
Kobayashi_2026_Imaging}.
Yet, the impact of spin textures on the electronic response in altermagnets remains
largely unexplored,
with only a few recent works addressing texture dynamics
and associated transport phenomena~\cite{Vakili_2025_Spin, Jin_2024_Skyrmion,
Jiang_2025_Altermagnet, Xiao_2025_Anomalous, Fu_2025_All, Liu_2026_Current}.

While collinear antiferromagnets and altermagnets both have zero net magnetization,
their electronic responses to spin textures can differ sharply.
In conventional
antiferromagnets, the Hamiltonian is invariant under N\'eel reversal ($\bm n \to -\bm
n$), implying that texture-induced responses are even in $\bm n$. In altermagnets, by
contrast, the electronic structure can distinguish the sign of $\bm{n}$, making certain
observables potentially sensitive to texture chirality and enabling emergent gauge-field
signatures~\cite{Vakili_2025_Spin, Gomonay_2024_Structure, Jin_2024_Skyrmion}.
These
effects are analogous in spirit to texture-induced Berry-phase phenomena in
ferromagnets~\cite{Bruno_2004_Topological}.

In this work, we develop an effective low-energy theory for itinerant electrons in
spin-textured collinear altermagnets using the SU(2) gauge
formalism~\cite{Tatara_2008_Microscopic, Tatara_2019_Effective, Davier_2023_Texture,
Davier_2024_Electron, Yokouchi_2020_Emergent} and show that gradients of the N\'eel
order act as emergent gauge fields on the electronic pseudospin.
This framework has three
generic consequences: it generates a texture-induced pseudospin-orbit coupling, yields an
emergent electromagnetic coupling in the presence of texture singularities, and produces
a texture-controlled pseudospin splitting.
We illustrate these effects for representative
$d$-wave and $g$-wave altermagnets and, for a coplanar spin helix, show that the helix
wave vector controls both the magnitude and the principal axes of the dc conductivity and
the optical absorption.
The altermagnetic channel is isolated by the momentum-resolved sublattice-odd
spectral structure and the interband dichroism, both of which vanish when $K^z=0$ and
neither of which requires intrinsic spin-orbit coupling.
We identify two response
regimes: at finite but low frequencies the dominant axis is locked to the crystal axes,
whereas at high frequencies it enters a tracking regime where it follows the helix
orientation.
Unlike spin-dependent transport~\cite{Song_2025_Altermagnets} and
valley-selective dichroism~\cite{Wang_2026_Valley} in \emph{homogeneous} altermagnets,
the texture-controlled changes identified here vanish in
the uniform crystal.

\paragraph*{Model---} A minimal description starts from collinear antiferromagnetic order
complemented by hopping processes that do not respect the translation that maps one
magnetic sublattice onto the other~\cite{Roig_2024_Minimal}.
In this setting, the
antiferromagnetic order enforces the antialignment of the local moments on the two
sublattices.
In this setting, a spin texture is described by a slowly varying N\'eel vector $\bm
n(\bm r)$, which defines the local spin quantization axis.  We do not model the
microscopic stabilization of the magnetic texture.
We instead take a smooth N\'eel
texture as a prescribed background and compute the electronic response to it.

The itinerant electronic Hamiltonian in the absence of intrinsic spin-orbit coupling (SOC) is
\begin{align}
  H &= H_\mathrm{kin}(\bm{k}) \sigma_0 - J\,\bm{n}(\bm{r}) \cdot \bm{\sigma}\,\eta_z,
  \label{eq:H_generic}
\end{align}
where $\bm{k} = (k_x, k_y)$ is the crystal momentum in a two-sublattice ($A,B$) unit
cell, $H_\mathrm{kin}(\bm{k})$ encodes spin-independent hopping processes,
$\eta_\alpha$ ($\sigma_i$) are Pauli matrices acting in sublattice (spin) space, and $J$
is the exchange coupling.

We consider a two-dimensional altermagnet and use a continuum parametrization around an
inversion-symmetric point,
\begin{equation}
  \begin{split}
    H_\mathrm{kin}(\bm k)=&\frac{k^2}{2m} +\left[C^x+\frac{K^x}{2}k^2\right]\eta_x
    +\frac{K^z}{n!}\,g_n(\bm k)\,\eta_z,
  \end{split}
  \label{eq:Hkin}
\end{equation}
where $m$ sets the isotropic dispersion, $C^x$ and $K^x$ the sublattice hybridization,
and $K^z$ is the symmetry-allowed staggered anisotropy associated with the basis function
$g_n(\bm{k})$, with $g_2(\bm{k})=k_x k_y$ for the $d$-wave case and $g_4(\bm{k})=k_x
k_y(k_x^2-k_y^2)$ for the $g$-wave case.

We treat an inhomogeneous N\'eel texture $\bm{n}(\bm{r})$  using a local unitary
$U(\bm{r})\in \mathrm{SU}(2)$ that aligns the spin quantization axis with the exchange
field~\cite{Tatara_2008_Microscopic,Tatara_2019_Effective,Davier_2023_Texture,Davier_2024_Electron,Yokouchi_2020_Emergent}
such that $ U^\dagger(\bm{r})\,\bm{n}(\bm{r})\cdot\bm{\sigma}\,U(\bm{r})=\sigma_z $.
Spatial gradients generate a connection
\begin{equation}
  D_j \equiv \partial_j + U^\dagger \partial_j U \equiv \partial_j -
  \frac{i}{2}\,\bm{A}_j\cdot\bm{\sigma},
\end{equation}
so in the comoving frame the exchange field becomes uniform, while the kinetic
energy couples to the texture through $\partial_j \to D_j$.
To handle operator ordering,
we define the gauge-coupled kinetic term by Weyl symmetrization as
\begin{equation}
  H = \mathcal{W}\!\left[H_\mathrm{kin}(k_j \to -iD_j)\right] - J\,\sigma_z \eta_z,
\end{equation}
where $\mathcal{W}[\cdot]$ denotes full symmetrization over noncommuting covariant derivatives.

The unitary $U(\bm r)$ defines a comoving orthonormal triad $(\bm e_1,\bm e_2,\bm{n})$
and allows one to decompose the connection into components transverse and longitudinal to $\bm n$,
\begin{equation}
  \bm A_j=\bm A_j^\perp + A_j^\parallel\,\bm n.
\end{equation}
The transverse component is uniquely fixed by the texture, $\bm A_j^\perp=\bm n\times
\partial_j \bm n$, whereas the longitudinal part $A_j^\parallel = -\,\bm e_1\cdot
\partial_j \bm e_2$ is the residual $\mathrm{U}(1)_z$ gauge field (rotations of $\bm
e_{1,2}$ about $\bm n$). Observables depend only on gauge-invariant combinations such as
the texture metric,
\begin{equation}
  g_{jk}\equiv \bm A_j^\perp\!\cdot \bm A_k^\perp = \partial_j \bm n\cdot \partial_k \bm n,
\end{equation}
as well as linear coupling in the transverse connection of the form $\bm A_j^\perp
\partial_k$ and, for textures with topological defects, the skyrmion density.

We focus on the $d$-wave case ($n=2$), while the general band expansion can be found in
Supplemental Material (SM)~\cite{SM}.
Including the texture through covariant
derivatives, the resulting Hamiltonian becomes
\begin{equation}
  \begin{split}
    H_d =\; &C^x \eta_x -\frac{1}{2m}\,\eta_0 D^2 -\frac{K^x}{2}\,\eta_x D^2 \\
    &-\frac{K^z}{2} \frac{\{D_x,D_y\}}{2}\eta_z - J\,\sigma_z \eta_z .
  \end{split}
\end{equation}
The Weyl symmetrized product reduces to an anticommutator and naturally splits into three
components:
\begin{equation}
  \begin{split}
    \label{eq:anticommutator}
    &\frac{1}{2}\{D_j,D_k\} = { \left(\partial_j-iA_j^\parallel\frac{\sigma_z}{2}\right)
    \left(\partial_k-iA_k^\parallel\frac{\sigma_z}{2}\right)} \\
    - &{i \left(\bm A_j^\perp\partial_k+\bm A_k^\perp\partial_j\right) \cdot
    \frac{\boldsymbol\sigma^\perp}{2}} -{\frac{g_{jk}}{4}\,\sigma_0}.
  \end{split}
\end{equation}
The first term represents the \emph{emergent electromagnetic coupling} to the Abelian
gauge field $A_j^\parallel$; the second term represents the \emph{emergent
pseudospin-orbit coupling} generated by the transverse texture, and the third term is a
\emph{scalar potential} proportional to the texture metric.

We work in the exchange-dominated regime, where the doublet separation $2E_0$ exceeds
the spin- and sublattice-dependent kinetic scales at the occupied Fermi surface and
treat those terms perturbatively.
Let $P=\tfrac{1}{2}\left(1+\sigma_z\eta_z \right)$ denote the projector
onto the low-energy subspace spanned by $\{|A \uparrow\rangle,\;|B \downarrow\rangle\}$.
By projecting into the low-energy sector, the resulting effective Hamiltonian reads as
\begin{equation}
  \label{eq:Hd_eff}
  \begin{split}
    &H_{d, \mathrm{eff}} = P H_{\mathrm{kin}} P =  \frac{1}{2m} \left[\sum_{\alpha=x, y}
    \left(- i D^\parallel_\alpha \right)^2 + \frac{g_{\alpha\alpha}}{4}  \tau_0 \right]\\
    &-  \frac{K^z}{2} \tau_3 \left[D^\parallel_x D^\parallel_y -
    \frac{ g_{xy}}{4} \tau_0 \right] - i {K^x} (\bm{A}^\perp_x
    \partial_x + \bm{A}^\perp_y \partial_y) \cdot \frac{\bm{\tau}}{2},
  \end{split}
\end{equation}
where $D^\parallel_\alpha = \partial_\alpha  - i {A}^\parallel_\alpha \tfrac{\tau_3}{2}$
and $\tau_i$  are the Pauli operators inside the two-component subspace.
A nonzero $C^x$ is treated nonperturbatively and renormalizes the coefficients~\cite{SM}.
Concretely, the expansion then requires only the low-energy hierarchy
$K^{x,z}k_F^2,\,v_Fq\ll2E_0$ and a Fermi energy below the gap $2E_0 = 2\sqrt{J^2+(C^x)^2}$,
with $k_F$ and $v_F$ being the Fermi momentum and velocity, respectively.

Equation~\eqref{eq:Hd_eff} yields a compact low-energy theory for electrons coupled to an
arbitrary smooth N\'eel texture in a collinear altermagnet.
For a conventional collinear
antiferromagnet ($K^{z}=0$), the sublattice-odd channel is absent and the effective
theory contains no $\tau_{3}$ term, so texture-induced responses remain sublattice even.
For an altermagnet ($K^{z}\neq 0$), the texture couples to the sublattice-odd sector and
produces a local $\tau_{3}$ component set by metric contractions of texture gradients,
i.e., an emergent pseudospin splitting.
Consequently, textures such as domain walls and
spirals can imprint a localized $\tau_{3}$ signal in observables like the
sublattice-resolved spectral function.

\paragraph*{Spin helix---}

\begin{figure}[ht]
  \centering
  \includegraphics[width=\columnwidth]{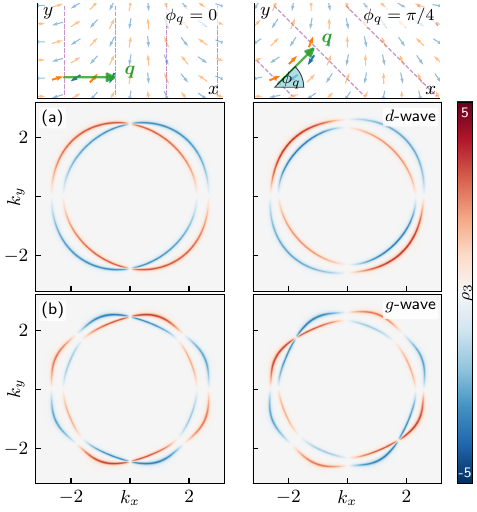}
  \caption{Real-space spin helices and their sublattice-resolved spectral imprint. Top:
    schematic planar antiferromagnetic helices with propagation vector $\bm
    q=q\cos(\phi_q)\,\hat{\bm x}+q\sin(\phi_q)\,\hat{\bm y}$ for $\phi_q=0$ (left) and
    $\pi/4$ (right), measured from the crystalline $\hat{\bm x}$ axis. Dashed lines are
    constant-phase lines perpendicular to $\bm q$. Rows (a) and (b) show the
    sublattice-odd spectral weight $\rho_3(\bm k)$ on the Fermi contours in the
    $(k_x,k_y)$ plane for a $d$-wave and $g$-wave altermagnet, respectively, at the same
    two helix orientations. The helix splits the Fermi surface along $\bm q$ and reorients
  the momentum-resolved sublattice-odd weight with $\phi_q$.} \label{fig:helix_FS}
\end{figure}

A coplanar spin helix, which captures the local rank-one constant-gradient limit of an
arbitrary smooth texture, is one of the simplest and most physically relevant spin
textures.
Its pitch and orientation may be set by Dzyaloshinskii--Moriya interactions,
strain gradients, boundary conditions, or other micromagnetic mechanisms that break
inversion symmetry~\cite{Bode_2007_Chiral}, which are not included explicitly in
Eqs.~\eqref{eq:H_generic} and \eqref{eq:Hkin}.
Consider a planar helix with propagation
wave vector $\bm q=(q_x,q_y)=q\,(\cos\phi_q,\sin\phi_q)$, whose orientation angle $\phi_q$
is measured from the crystalline $\hat{\bm x}$ axis, parametrized as
\begin{equation}
  \bm n(\bm r)=\bm u \cos(\bm q \cdot \bm r)+\bm v \sin(\bm q \cdot \bm r),
\end{equation}
where $\bm u$ and $\bm v$ are fixed orthonormal unit vectors spanning the helix plane
and $\bm w\equiv \bm u\times \bm v$ is the normal to that plane, such that $\bm n\times
\partial_i \bm n=q_i\,\bm w$ [top panels of Fig.~\ref{fig:helix_FS}]. Choosing the
comoving frame with $\bm e_1=\bm w$ fixes $A_j^\parallel=0$ and $\bm A_j^\perp=q_j\,\bm e_1$.

In the $d$-wave case, the long-wavelength Hamiltonian for itinerant electrons reads as
\begin{equation}
  \label{eq:Hd_helix}
  \begin{split}
    &H_{d, h} =\frac{1}{2m} \left[k_x^2+k_y^2+\frac{q_x^2+q_y^2}{4}\right] \\
    &+\,\frac{K^x \,\tau_1}{2}\,\left[q_x k_x + q_y k_y\right] +\frac{K^z\,\tau_3}{2}
    \left[k_x k_y+\frac{q_x q_y}{4}\right].
  \end{split}
\end{equation}
The analogous $g$-wave Hamiltonian, obtained at quartic order in gradients, has the same
$\tau_1$ texture coupling and a $\tau_3$ channel that mixes lower harmonics; its explicit
form is given in SM~\cite{SM}.

Within the effective theory, a smooth helix generates two texture-induced effects.
First,
the emergent pseudospin-orbit coupling induces a Fermi surface splitting.
Second, it modifies the momentum-resolved sublattice-odd spectral weight
$\rho_3(\bm k,\omega) \equiv
(2\pi)^{-1}\,\mathrm{Im}\,\mathrm{Tr}\!\left[\tau_3\,(G^{A}-G^{R})\right]$.
Its momentum integral defines the net polarization
$\mathcal P_3(\omega)\equiv\int_{\bm k}\rho_3(\bm k,\omega)$.
Here
$G^{R,A}(\bm k,\omega)=[\omega+\mu-H(\bm k)\pm i\Gamma]^{-1}$ are Green's functions,
$\mu$ is the chemical potential, and $\Gamma$ is an effective broadening.
The key distinction between the $d$-wave and $g$-wave models is the momentum structure of
this $\tau_3$ contribution: in the $g$-wave case, the contraction of the intrinsic
$\ell=4$ anisotropy with the texture metric lowers the effective symmetry, making the
induced $\tau_3$ polarization alternate in sign around the contour.
For helix directions
along crystal axes, the continuum model admits symmetry-enforced degeneracy lines that
intersect a given Fermi contour at isolated points; away from these special orientations,
the helix generically lifts the degeneracy and splits the Fermi surfaces anisotropically
[Figs.~\ref{fig:helix_FS}(a) and \ref{fig:helix_FS}(b)].


The texture-induced $\tau_3$ term acts as an emergent pseudospin Zeeman field controlled
jointly by the altermagnetic form factor $K^z$ and the texture metric; it therefore
vanishes in either the conventional-antiferromagnetic ($K^z=0$) or uniform limit.
The net polarization $\mathcal P_3$ vanishes as $q\to0$, although the momentum-resolved
$\rho_3(\bm k,\omega)$ need not.
This momentum-resolved structure can be probed by atomically registered Fourier-transform scanning
tunneling spectroscopy: subtracting the quasiparticle-interference maps of the two
magnetic sublattices isolates the odd scattering
channel~\cite{Gondolf_2025_Local,Petermann_2025_Spin}.


For a homogeneous N\'eel state ($q=0$), tetragonal symmetry enforces
$\sigma_{xx}=\sigma_{yy}$.
A spin helix lowers the symmetry and unlocks an anisotropic
correction.
It is therefore natural to resolve the dc conductivity in the spiral basis
$(\hat{\bm q},\hat{\bm q}_\perp)$, with
$\sigma_{\parallel}=\hat{\bm q}\cdot\boldsymbol\sigma^{\mathrm{dc}}\cdot\hat{\bm q}$,
$\sigma_{\perp}=\hat{\bm q}_{\perp}\cdot\boldsymbol\sigma^{\mathrm{dc}}\cdot\hat{\bm q}_{\perp}$,
and
$\sigma_{\times}=\hat{\bm q}\cdot\boldsymbol\sigma^{\mathrm{dc}}\cdot\hat{\bm q}_{\perp}$.
The splitting $\sigma_{\parallel}-\sigma_{\perp}$ measures the helix-induced anisotropy,
while $\sigma_{\times}$ captures any residual misalignment between the helix and the
conductivity principal axes.
Altermagnetic helices exhibit
$\sigma_{\parallel}\neq\sigma_{\perp}$, with a pronounced tetragonal modulation in the
$d$-wave case and a weaker, $\pi/4$-periodic modulation in the $g$-wave
case~\cite{SM}.
In both cases,
$\sigma_{\times}$ remains subleading, indicating that $\hat{\bm q}$ closely tracks the
principal axes of the symmetric conductivity tensor.

This anisotropy is genuine but not, by itself, unique to altermagnetism:
a conventional antiferromagnetic helix
exhibits it as well.
In the matched isotropic conventional-AFM model,
$\sigma_{\parallel}-\sigma_{\perp}$ is independent of the helix orientation and
$\sigma_{\times}=0$, so the tensor rigidly follows the helix.
The altermagnet is
quantitatively distinguished by an anisotropy growing as $\mathcal{O}(q)$ rather than
$\mathcal{O}(q^2)$~\cite{Bonbien_2020_Symmetrized, SM}.
Reproducing its crystal-locked
modulation and finite $\sigma_{\times}$ in a conventional AFM requires an additional
crystal-locked $C_4$ anisotropy~\cite{SM}.
The dc angular structure is therefore a
quantitative rather than absolute discriminator.
The qualitative altermagnetic
fingerprint instead lies in the interband optical response, to which we now turn.


Turning to optical absorption, we find that the system exhibits \emph{linear dichroism}:
the dissipated power depends on the in-plane polarization direction, and two orthogonal
linear polarizations are absorbed differently at the same frequency.
In linear response
this is encoded in the dissipative optical conductivity,
$\mathrm{Re}\,\sigma_{ij}(\omega)$, whose two eigenvalues
$\sigma_{+}(\omega)\neq\sigma_{-}(\omega)$ correspond to absorption along orthogonal
in-plane principal axes.
We quantify the dichroism strength by
\begin{equation}
  \mathfrak
  D(\omega)\equiv\frac{\sigma_{+}(\omega)-\sigma_{-}(\omega)}{\sigma_{+}(\omega)+\sigma_{-}(\omega)}
\end{equation}
and introduce the polarization-averaged response $\overline{\sigma}(\omega)\equiv
[\sigma_{+}(\omega)+\sigma_{-}(\omega)]/2$, which sets the absorption for unpolarized
light.
The principal absorption axis $\theta_{+}(\omega)$ is defined as the eigenvector
angle associated with $\sigma_{+}(\omega)$.

The optical conductivity is obtained from the standard relation $
\sigma_{ij}(\omega)=(i\omega)^{-1}[K^R_{ij}(\omega)- K^R_{ij}(0)], $ where
$K^R_{ij}(\omega)$ is the retarded current-current correlator.
To analyze anisotropy and
symmetry, we focus on the interband paramagnetic contribution,
\begin{equation}
  \begin{split}
    K^{\mathrm{para}}_{ij}(\omega)= -e^{2}\!\int\! d\bm{k}\sum_{n\neq m}
    \frac{f_{n\bm{k}}-f_{m\bm{k}}} {\omega+i\Gamma-\Delta_{mn}}\, v^{nm}_{i}v^{mn}_{j},
  \end{split}
  \label{eq:Kpara}
\end{equation}
where $v^{nm}_{i}(\bm k)$ are the interband velocity matrix elements encoding the full
angular dependence, $\Delta_{mn} = E_{m\bm{k}}-E_{n\bm{k}}$, and
$f_{n\bm{k}}=\bigl[e^{E_{n\bm{k}}/T}+1\bigr]^{-1}$ are Fermi-Dirac occupation factors at
temperature $T$.

\begin{figure}[ht]
  \centering
  \includegraphics[width=\linewidth]{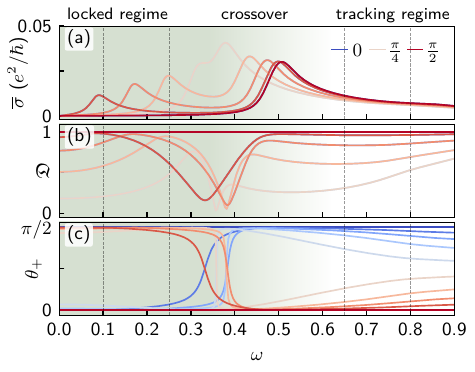}
  \caption{
    (a)~Total optical absorption $\overline{\sigma}(\omega)$ for ten helix orientations
    $\phi_q$ from $0$ (blue) to $\pi/2$ (red), color coded as in the legend
    ($0,\pi/4,\pi/2$ marked).
    Reflection about the $k_x=k_y$ axis maps
    $\phi_q\to\pi/2-\phi_q$ and leaves $\overline{\sigma}$ and $\mathfrak D$ invariant,
    so the curves at $\phi_q$ and $\pi/2-\phi_q$ coincide in (a) and (b); in particular
    the $\phi_q=0$ (blue) and $\pi/2$ (red) lines overlap.
    (b)~Corresponding linear
    dichroism $\mathfrak D(\omega)$.
    (c)~Frequency-dependent orientation
    $\theta_{+}(\omega)$ of the principal absorption axis.
    The vertical dashed lines mark
    the four probe frequencies used in Fig.~\ref{fig:absorption_axis}(a).
  }
  \label{fig:dichroism}
\end{figure}

Since the helix breaks the crystalline $C_{4}$ symmetry down to $C_{2}$, the tensor
structure of $K^{\mathrm{para}}_{ij}(\omega)$ becomes angle dependent.
Specifically, the
interband absorption is controlled by the off-diagonal velocity matrix elements that,
for the $d$-wave helix with $q\ll k_F$, give
\begin{equation}
  \label{eq:vpm_dwave}
  v_i^{+-}\approx-\frac{K^x K^z}{4}\, \frac{q_{\bar i}\,k_{\bar i}^{2}}{\sqrt{d_1^2+d_3^2}},
\end{equation}
where $(\bar x\equiv y,\ \bar y\equiv x)$; see SM~\cite{SM}.
Equation~\eqref{eq:vpm_dwave} shows that
the optical anisotropy is not trivially locked to $\bm q$ since $x$-polarized transitions
are controlled by $q_y$ (and vice versa), reflecting interference between the
helix-induced mixing through pseudospin-orbit coupling and the crystal form factor.

In conventional collinear antiferromagnets, linear dichroism is usually attributed to
intrinsic spin-orbit or crystal-field anisotropies~\cite{Nemec_2018_Antiferromagnetic,
Grigorev_2021_Optical}.
In their absence it is symmetry forbidden: for translation-connected sublattices and a coplanar
texture, a $\pi$ spin rotation about $\bm w$ composed with the sublattice-connecting translation
commutes with $H$ and with the current, and the two bands are its $\pm1$ eigenstates, so
$v_i^{+-}\equiv0$~\cite{SM, Smejkal_2022_Beyond}.
By contrast, in an
altermagnet, a distinct \emph{texture-induced interband mechanism} is already present
without intrinsic spin-orbit coupling since the textured altermagnet features a
nonrelativistic spin-split structure with momentum-dependent eigenvectors that feed
directly into interband optical matrix elements.
It is also distinct from the
valley-selective dichroism of \emph{homogeneous} altermagnets~\cite{Wang_2026_Valley}, a
spin-conserving selection rule that yields no net dichroism unless the fourfold symmetry
is broken externally; here the texture supplies both the symmetry breaking and the
interband coupling, giving a net dichroism whose axes track the helix.

\begin{figure}[ht]
  \centering
  \includegraphics[width=\linewidth]{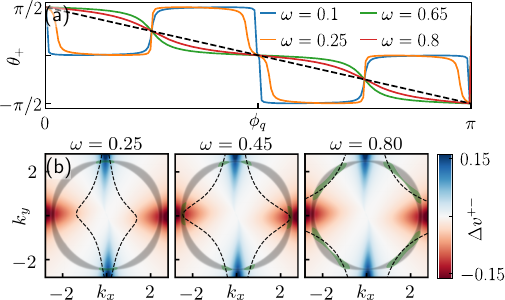}
  \caption{(a) Principal absorption-axis angle $\theta_+(\omega)$ versus helix
    orientation $\phi_q$. In the \emph{locked} regime $\theta_+$ is pinned to the crystal
    axes (plateaus with $\pi/2$ jumps), while in the \emph{tracking} regime it varies
    smoothly and approaches $\theta_+\simeq \pi/2-\phi_q$ (dashed). (b)
    Interband-velocity anisotropy $\Delta v^{+-}\equiv |v_x^{+-}|^2-|v_y^{+-}|^2$ for
    the three indicated probe frequencies at fixed helix orientation $\phi_q=\pi/8$. Black dashed:
    resonance $E_{+}(\bm k)-E_{-}(\bm k)=\omega$. Thick gray: Fermi surface. Green
    shading: absorption weight $W_\omega(\bm k)$ (Pauli factor times resonance kernel),
    concentrated where the resonance contour crosses the Pauli-allowed region near the
    Fermi surface.
  }
  \label{fig:absorption_axis}
\end{figure}

The overall line shape of  $\overline{\sigma}(\omega)$ is controlled by the interband
resonance condition, while the relative weight of its features depends on the helix
orientation $\phi_q$ [Figs.~\ref{fig:dichroism}(a) and \ref{fig:dichroism}(b)].
This angular dependence reflects
that the relevant interband matrix elements are governed by the texture-induced mixing
and its interference with the altermagnetic form factor.
Two regimes separated by a
crossover frequency $\omega_c(\phi_q)\equiv\operatorname*{arg\,min}_{\omega}\mathfrak
D(\omega)$ can be identified. For $\omega\lesssim\omega_c$, the response is strongly
polarization selective and $\mathfrak D(\omega)\approx 1$ over a broad window, indicating
absorption dominated by a single linear polarization [Fig.~\ref{fig:dichroism}(b)].
In
this \emph{locked regime}, the principal axis $\theta_+$ develops broad pinning plateaus
and undergoes sharp $\pi/2$ reorientations as $\phi_q$ is varied
[Fig.~\ref{fig:absorption_axis}(a)].
Near $\omega_c$, $\mathfrak D(\omega)$ shows a
pronounced dip accompanied by an abrupt $\pi/2$ rotation of $\theta_+$, signaling a
narrow \emph{crossover window} where the two eigenvalues of $\mathrm{Re}\, \sigma$ become
nearly degenerate.
For $\omega\gtrsim\omega_c$ the response crosses over to a
\emph{tracking regime} in which $\theta_+(\omega)$ varies smoothly and approaches a
helix-tracking law,
\begin{equation}
  \theta_+ \simeq \frac{\pi}{2}-\phi_q .
  \label{eq:tracking_law_dwave}
\end{equation}
Equation~\eqref{eq:tracking_law_dwave} implies that, in the tracking regime, the absorptive
eigenbasis is not the spiral frame $(\hat{\bm q},\hat{\bm q}_\perp)$ but approaches the
swapped direction $\hat{\bm e}_{\mathrm{tr}}=(\sin\phi_q,\cos\phi_q)$, i.e., the helix
direction reflected by $x\leftrightarrow y$.


This behavior can be understood by expressing the absorptive tensor in terms of interband matrix
elements.
Schematically, the interband contribution has the form
\begin{equation}
  \label{eq:M_vpm}
  \mathrm{Re}\, \sigma_{ij}(\omega) \propto \int d\bm{k}\;
  v_i^{-+}(\bm{k})\,v_j^{+-}(\bm{k})\; W_\omega(\bm{k}),
\end{equation}
where $W_\omega(\bm{k})$ is a peaked function centered on hot spots placed along the resonance
condition $\omega=E_+(\bm{k})-E_-(\bm{k})$, with a width set by $\Gamma$ and $T$.
The interference
between the twofold texture gap and the fourfold altermagnetic gap produces two inequivalent
minima in the Fermi-surface splitting.
Deviations for $\omega\lesssim\omega_c$ arise when
$W_\omega(\bm{k})$ is dominated by anisotropic hot spots along the resonant contour, so the
resonance average is no longer approximately $x\leftrightarrow y$ symmetric.
In particular,
for $\omega\ll \omega_c$ only the lower valley is accessible, and the weight concentrates near the
crystal axes ($k_x\simeq 0$ or $k_y\simeq 0$), causing the switching regime.
When the second
valley becomes accessible, additional resonant states with a competing polarization enter and
drive the dichroism dip, placing $\omega_c$ near this threshold.
The resulting scaling form is
$\omega_c\simeq\tfrac{1}{2}K^zk_F^2\,X(\rho,\phi_q)$, with $\rho=2 K^x q/(K^z k_F)$.
For finite weak
textures ($\rho\ll1$), $\omega_c\sim K^x q k_F$ up to an orientation
factor~\cite{SM}.
For band parameters representative of layered $d$-wave
altermagnets and texture scales already imaged in altermagnets, the locked-to-tracking
dichroism spans the terahertz-to-mid-infrared window~\cite{SM}.

At high $\omega$, the resonant weight samples the contour almost uniformly, so the resonance average
becomes approximately $x\leftrightarrow y$ symmetric [Fig.~\ref{fig:absorption_axis}(b)].
In this
regime, using Eq.~\eqref{eq:vpm_dwave}, we can rewrite Eq.~\eqref{eq:M_vpm} as
\begin{align}
  \label{eq:M_entries}
  \mathrm{Re}\, \sigma_{ij}(\omega) &\propto
  q_{\bar i}q_{\bar j} \left\langle \frac{k_{\bar i}^2 k_{\bar
  j}^2}{d_1^2+d_3^2}\right\rangle_{W_\omega},
\end{align}
and $x\leftrightarrow y$ symmetric sampling implies the factorization $\mathrm{Re}\, \sigma\propto
(q_y,q_x)(q_y,q_x)^T$, resulting in the tracking law in Eq.~\eqref{eq:tracking_law_dwave}.

\paragraph*{Discussion---} N\'eel textures provide a direct way to distinguish altermagnets
from conventional collinear antiferromagnets in electronic response.
The texture induces
three generic ingredients: an emergent U(1) gauge coupling tied to topological defects, a
metric-controlled pseudospin splitting, and a transverse texture-induced pseudospin-orbit coupling.

For a coplanar spin helix, the altermagnetic contribution is isolated by two fingerprints
that vanish when $K^z=0$: the momentum-resolved sublattice-odd spectral structure and the SOC-free
interband linear dichroism, a more readily accessible optical probe of the same channel.
Interband absorption becomes polarization selective and undergoes a characteristic
frequency-dependent reorientation, producing linear dichroism that tracks the helix
orientation.
Together, sublattice-sensitive spectroscopy and polarization-resolved optics
provide direct probes of textured altermagnetic states and a practical discriminator from
collinear antiferromagnets.
As shown in the End Matter, these signatures survive weak
intrinsic spin-orbit coupling, which does not spoil the SOC-free mechanism but instead renders the
helix-plane normal, and hence the texture chirality, optically observable.

Additionally, the dc conductivity becomes anisotropic, but this is a quantitative rather
than qualitative discriminator as related antiferromagnetic helices can also exhibit
anisotropic transport.

The framework is independent of microscopic details and extends naturally to other
altermagnetic symmetries via the appropriate form factors.
It can incorporate
higher-gradient corrections and time-dependent textures, and it provides a route to
texture-engineered phenomena, including defect-pinned bound
states~\cite{Davier_2023_Texture, Davier_2024_Electron} and spatially programmable
anisotropies relevant for spintronics~\cite{Liu_2025_Altermagnetic}.
Combined with
superconducting proximity~\cite{Fukaya_2025_Superconducting, Maeda_2025_Classification},
it enables texture-controlled Andreev spectra~\cite{Pershoguba_2016_Skyrmion,
  Beenakker_2023_Phase, Sun_2023_Andreev, Maiani_2025_Impurity, Lu_2025_Engineering,
Fukaya_2025_Crossed} as well as magnetoelectric~\cite{Rabinovich_2019_Magnetoelectric}
and optoelectronic~\cite{Fu_2025_Light, Fu_2026_Floquet, Yarmohammadi_2026_Spin} responses.

\paragraph*{Note added---} Recently, we have become aware of an independent preprint
developing a closely related low-energy description of electrons in
altermagnetic textures~\cite{Schrade_2026_Altermagnetic}.
The results are complementary
to this work.

\paragraph*{Acknowledgments---} This work is funded by the Wallenberg Initiative on
Networks and Quantum Information.
A. M. thanks Mikael Fogelström, Alberto Cortijo, Ruben S.
Souto, and Stavros Komineas for useful discussions.

\paragraph*{Data availability---} The data that support the findings of this article are
openly available~\cite{Zenodo_repo}.

\nocite{Cheng_2026_Realistic}
\bibliography{am_textures}

\paragraph*{End Matter} The Hamiltonian~\eqref{eq:H_generic} omits intrinsic
(relativistic) spin-orbit coupling.
Since a spin helix is frequently stabilized by the
Dzyaloshinskii--Moriya interaction, which itself originates in SOC, it is important to
establish that weak intrinsic SOC does not spoil the texture-generated response and to
identify how it modifies it.
A full account is given in SM; here we summarize the result.

To linear order in $\bm k$, and once inversion is broken, invariance under time reversal,
the altermagnetic antiunitary $[\mathcal T C_{4z}\eta_x]$, and the
mirror-plus-sublattice-swap $M_x\eta_x$ leave three SOC channels~\cite{SM, AbadilloUriel_2026_All},
\begin{equation}
  \begin{split}
    H_{\mathrm{SOC}}=&\lambda_A\,\eta_0(\bm k\times\bm\sigma)_z +\lambda_B\,\eta_x(\bm
    k\times\bm\sigma)_z \\
    +&\lambda_C\,\eta_y\bigl[\bm k\times(\bm n\times\bm\sigma)\bigr]_z , \label{eq:em_hsoc}
  \end{split}
\end{equation}
two N\'eel-even Rashba terms ($\lambda_{A,B}$) and one N\'eel-odd, texture-locked term
($\lambda_C\propto J$).
Projecting onto the low-energy doublet intrinsic SOC does not
open any new operator at first order: it enters \emph{only} as an additive shift of the
same emergent gauge fields generated by the texture.
For $C^x=0$,
Channels B and C fuse into a single transverse shift $\propto\bar\alpha=\lambda_B+\lambda_C$,
while the uniform Rashba $\lambda_A$ shifts the Abelian field.
Explicitly, projecting Eq.~\eqref{eq:em_hsoc} onto
the doublet shifts the two emergent connections of the texture,
\begin{equation}
  \delta\!\bm A_j^{\perp}=-\frac{2\bar\alpha}{K^x}\bigl(\hat{\bm z}\times\hat{\bm
  x}_j\bigr)_{\perp\bm n}, \quad \delta\!A_j^{\parallel}=-2m\lambda_A\bigl(\bm
  n\times\hat{\bm z}\bigr)_j , \label{eq:em_soc_shift}
\end{equation}
where $\{\hat{\bm x}, \hat{\bm y}, \hat{\bm z}\}$ are the lab-frame axes, $(\cdot)_{\perp\bm n}$
denotes the component transverse to $\bm n$, and the Abelian shift carries the emergent
magnetic field $\delta b=2m\lambda_A\,\nabla\!\cdot\bm n_\parallel$~\cite{SM}.
Unlike the
texture connections, whose direction is set by the texture and whose magnitude is
$\propto \partial\bm n$, the SOC shift is crystal pinned and depends on $\bm{n}$ itself.

For a helix with plane normal $\bm w=\bm u\times\bm v$, averaging
Eq.~\eqref{eq:em_soc_shift} over the helix phase yields a rigid renormalization of the
helix wave vector in the $\tau_1$ channel,
\begin{equation}
  \bm q_{\mathrm{eff}}=\bm q+\underbrace{\frac{2\bar\alpha}{K^x}\,(\hat{\bm z}\times\bm
  w)}_{\delta \bm{q}}, \qquad \zeta\equiv\frac{2\bar\alpha}{K^xq}\,\bigl|\bm
  w\times\hat{\bm z}\bigr| , \label{eq:em_qeff}
\end{equation}
with $\zeta$ being the dimensionless ratio of the SOC shift to the texture gauge field.
The
effective theory is thus \emph{form-invariant}: the entire SOC-free machinery applies
with $\bm q\to\bm q_{\mathrm{eff}}$, and the locked-to-tracking crossover survives.

\begin{figure}[ht]
  \centering
  \includegraphics[width=\columnwidth]{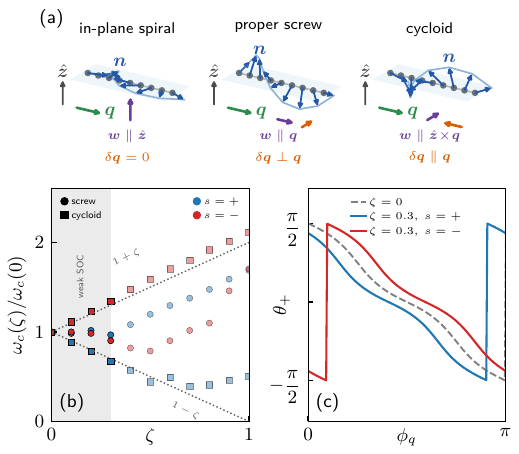}
  \caption{Effect of intrinsic SOC on the optical crossover, parametrized by
    $\zeta$ [Eq.~\eqref{eq:em_qeff}]; blue and red denote the two helix chiralities $s=\pm$
    ($\bm w\to-\bm w$). (a)~Representative in-plane spiral, proper screw, and cycloid in a
    quasi-two-dimensional interface. Blue arrows show the N\'eel vector $\bm n$; green, purple,
    and orange arrows show $\bm q$, $\bm w$, and $\delta\bm q$, respectively.
    (b)~Numerically exact crossover frequency
    $\omega_c(\zeta)/\omega_c(0)$, identified by the dichroism minimum that carries the
    locked-to-tracking principal-axis switch, from a superlattice Bloch-ladder calculation
    of the four-band helix model with all winding terms retained, for the screw (circles)
    and cycloid (squares). Dotted lines are the cycloid guides $1\pm\zeta$. (c)~Principal
    absorption axis $\theta_+(\phi_q)$ in the tracking regime from the projected model, for
    the screw at $\zeta=0$ (dashed) and $\zeta=0.3$ (solid): SOC shifts the whole tracking
  law rigidly by the chirality-odd angle $\mp\arctan\zeta$.} \label{fig:soc}
\end{figure}

Intrinsic SOC therefore makes the helix-plane normal $\bm w$ observable.
In the SOC-free
theory this orientation can be changed by a global spin rotation, but
Eq.~\eqref{eq:em_qeff} couples $\bm w$ to the crystal axis $\hat{\bm z}$ through
$\delta\bm q$.
This gives three simple limits.
For an in-plane spiral ($\bm
w\parallel\hat{\bm z}$), $\delta\bm q=0$, so the first-order SOC correction vanishes. For
a proper screw with chirality $s=\pm1$ ($\bm w=s\hat{\bm q}$), $\delta\bm q\perp\bm q$ and
$\bm q_{\mathrm{eff}}/q=\hat{\bm q}+s\zeta(\hat{\bm z}\times\hat{\bm q})$, so its azimuth is
$\phi_{\mathrm{eff}}=\phi_q+\arctan(s\zeta)$.
Since the tracking law becomes
$\theta_+\simeq\pi/2-\phi_{\mathrm{eff}}$, the absorption axis shifts by the chirality-odd
angle $\delta\theta_+=-\arctan(s\zeta)=\mp\arctan\zeta$.
For a cycloid ($\bm
w\perp\hat{\bm q}$), $\delta\bm q\parallel\bm q$ and
$q_{\mathrm{eff}}=(1\mp\zeta)q$, so $\omega_c\to(1\mp\zeta)\,\omega_c$
[Fig.~\ref{fig:soc}(b)], as derived in SM~\cite{SM}.
Since $\delta\bm q$ changes sign
under helix-chirality reversal $\bm w\to-\bm w$, intrinsic SOC makes the optical response
sensitive to the texture handedness selected by the stabilizing interaction, a
sensitivity absent in the SOC-free theory.

These first-order conclusions require $\zeta\ll1$; the projection additionally requires
$|\lambda_i|k_F\ll2E_0$ for $i=A,B,C$.
An exact four-band calculation retaining all
winding terms confirms the locked-to-tracking switch and its chirality-dependent
displacement in this regime [Fig.~\ref{fig:soc}(b); see SM~\cite{SM}].
\end{document}


\title{Emergent gauge fields and linear dichroism in spin-textured altermagnets:
Supplemental Material}

\author{Andrea Maiani}
\affiliation{Nordita, KTH Royal Institute of Technology and Stockholm University,
Hannes Alfvéns väg 12, SE-10691 Stockholm, Sweden}

\date{20 July 2026}
\maketitle

\tableofcontents

\section{SU(2) texture gauge formalism}
\label{app:su2_gauge_formalism}

In this section we briefly review the SU(2) gauge formulation commonly used to describe electrons
coupled to smooth spin textures.
We consider a translationally invariant spin-degenerate kinetic
Hamiltonian $H_\mathrm{kin}(\bm{k})$ supplemented by an exchange coupling to a (slowly varying)
unit vector texture $\bm{n}(\bm{r})$,
\begin{equation}
  H = H_\mathrm{kin} - J\, \bm{n}(\bm{r}) \cdot \bm{\sigma}.
\end{equation}
To treat this system, it is convenient to perform a local SU(2) rotation that aligns the spin
quantization axis with the exchange field.
Concretely, we introduce a local right-handed orthonormal
triad $(\bm{e}_1,\bm{e}_2,\bm{e}_3)$ with
\begin{equation}
  \bm{e}_3 \equiv \bm{e}_1\times \bm{e}_2 = \bm{n}.
\end{equation}
The choice of $(\bm{e}_1,\bm{e}_2)$ is not unique: a local rotation about $\bm{n}$ leaves $\bm{e}_3$
invariant and corresponds to a residual $\mathrm{U(1)}_z$ gauge freedom.
More explicitly, a
local rotation about $\bm{n}$ acts on the comoving frame as
\begin{equation}
  U(\bm{r}) \;\to\; U(\bm{r})\, e^{-i\chi(\bm{r})\sigma_3/2},
\end{equation}
with an arbitrary smooth function $\chi(\bm{r})$.
Under this transformation, the longitudinal
component of the gauge field shifts as
\begin{equation}
  A_j^\parallel \;\to\; A_j^\parallel + \partial_j \chi,
\end{equation}
while the transverse component $\bm{A}_j^\perp$ is unchanged.
Thus, $A_j^\parallel$ plays the
role of an emergent Abelian gauge potential, whereas $\bm{A}_j^\perp$ is fixed geometrically
by the texture itself.

The change of basis between the laboratory frame $(\bm{e}_x,\bm{e}_y,\bm{e}_z)$ and the comoving
frame $(\bm{e}_1,\bm{e}_2,\bm{n})$ is implemented by a unitary matrix $U(\bm{r})$ defined by
\begin{equation}
  \label{eq:unitary_definition}
  U^\dagger(\bm{r})\,\bm{n}(\bm{r})\cdot\bm{\sigma}\,U(\bm{r})=\sigma_3.
\end{equation}
The associated $\mathrm{SO(3)}$ rotation is
\begin{equation}
  \label{eq:rotation_from_U}
  R_{ij}(\bm{r}) = \frac{1}{2}\,\tr \!\left[ \sigma_i\, U(\bm{r})\, \sigma_j\,
  U^\dagger(\bm{r}) \right] = (\bm{e}_j)_i,
\end{equation}
so that $U \sigma_j U^\dagger = R_{ij}\sigma_i$ and the columns of $R$ are the triad vectors
in the laboratory basis, $R(\bm{r}) =
\begin{pmatrix} \bm{e}_1(\bm{r}) & \bm{e}_2(\bm{r}) & \bm{e}_3(\bm{r})
\end{pmatrix}$.

We define the non-Abelian gauge connection
\begin{equation}
  \mathcal{A}_j(\bm{r}) = i\, U^\dagger(\bm{r})\, \partial_j U(\bm{r})
  = \frac{1}{2}\,\bm{A}_j(\bm{r})\cdot \bm{\sigma},
\end{equation}
where $\bm{A}_j$ is a three-component vector expressed in the comoving frame.
The corresponding
vector expressed in the laboratory frame is $\tilde{\bm{A}}_j = R^{-1}\bm{A}_j$ (equivalently,
$\bm{A}_j = R\,\tilde{\bm{A}}_j$).

Working in the comoving frame where $R\,\bm{n}=\bm{e}_3$, we may write $\bm{n}=R^{-1}\bm{e}_3$
and obtain
\begin{equation}
  \label{eq:dn_dj}
  R\,\partial_j \bm{n}
  = R\,\partial_j \!\left(R^{-1} \bm{e}_3\right)
  = R\,(\partial_j R^{-1})\,\bm{e}_3
  = \bm{A}_j \times \bm{n}.
\end{equation}
Taking the cross product with $\bm{n}$ yields the key identity
\begin{equation}
  \bm{A}^\perp_j \equiv \bm{A}_j - (\bm{A}_j \cdot \bm{n})\, \bm{n}
  = \bm{n} \times \partial_j \bm{n},
\end{equation}
showing that the transverse component of the connection is completely fixed by the texture.

The longitudinal component is instead controlled by the $\mathrm{U(1)}_z$ gauge choice.
Using
that \eqref{eq:dn_dj} applies to any comoving basis vector (e.g. $\partial_j \bm{e}_2 = \bm{A}_j
\times \bm{e}_2$), we find
\begin{equation}
  A^\parallel_j \equiv \bm{A}_j \cdot \bm{n} = -\,\bm{e}_1 \cdot \partial_j \bm{e}_2.
\end{equation}
Since $(\bm{e}_1,\bm{e}_2)$ may be rotated locally about $\bm{n}$, $A^\parallel_j$ can be shifted
by a gradient and is therefore not itself gauge invariant.
Physical observables must be built
from gauge-invariant combinations of $\bm{n}$ and its derivatives.
A basic example is
\begin{equation}
  \bm{A}^\perp_j \cdot \bm{A}^\perp_k
  = \partial_j \bm{n} \cdot \partial_k \bm{n}
  \equiv g_{jk},
\end{equation}
which is the texture metric tensor associated with spatial variations of $\bm{n}$.

Under the local rotation, spatial derivatives become covariant, $\partial_j \to D_j \equiv
\partial_j - i\mathcal{A}_j$, so that in the comoving frame the kinetic part is obtained by the
minimal substitution $\bm{k}\to\bm\pi$, where $\bm\pi\equiv -i\bm D=-i\bm\nabla-\boldsymbol{\mathcal
A}$ is the gauge-covariant kinetic momentum.

Although the SU(2) gauge field is locally pure gauge, its curvature
\begin{equation}
  \mathcal{F}_{jk}=\partial_j\mathcal A_k-\partial_k\mathcal A_j-i[\mathcal A_j,\mathcal A_k]
\end{equation}
vanishes wherever $U(\bm r)$ is smooth.
Since $[D_j,D_k]=i\mathcal F_{jk}$, the covariant
derivatives commute pointwise, and any nonzero field strength signals a gauge singularity associated
with topological texture defects.

Projecting onto a fixed spin sector produces an emergent Abelian gauge field
\begin{equation}
  a_j \equiv \frac{1}{2}\,A^\parallel_j,
\end{equation}
with field strength
\begin{equation}
  f_{jk}=\partial_j a_k-\partial_k a_j
  =\frac{1}{2}\left(\partial_jA_k^\parallel-\partial_kA_j^\parallel\right)
  =2\pi\,\epsilon_{jk}\,\rho_s(\bm r),
\end{equation}
where $\epsilon_{xy}=+1$ and the skyrmion density is
\begin{equation}
  \rho_s(\bm r)=\frac{1}{4\pi}\,\bm n\cdot\left(\partial_x\bm n\times\partial_y\bm n\right)
  =\frac{1}{4\pi}\,\bm n\cdot\left(\bm A_x^\perp\times\bm A_y^\perp\right).
\end{equation}
Hence, the emergent flux $\int d^2r\, f_{xy}=2\pi\,Q$ is quantized by the skyrmion number $Q=\int
d^2r\,\rho_s$.

\section{Covariant expansion of inversion-symmetric band structures}
\label{sec:bandexpansion}
In this section, we derive the covariant long-wavelength expansion of an
inversion-symmetric kinetic Hamiltonian, keeping terms up to quartic order in
derivatives. We expand the kinetic Hamiltonian around an inversion-invariant momentum
$\bm{k}_0$ in the Brillouin zone. Around such points, inversion symmetry implies that the
Bloch Hamiltonian is an even function of momentum. In the comoving frame, the Hamiltonian reads
\begin{equation}
  \begin{split}
    H = &C^{x}\eta_x -J\sigma_3\eta_z + \sum_{\alpha\in\{0,x,z\}}
    \sum_{\substack{n=2\\\text{even}}}^{n_{\max}} \frac{(-i)^n}{n!}\, C^{\alpha}_{a_1
    \dots a_n}\, \eta_{\alpha}\, \mathcal W \bigl[D_{a_1} \dots D_{a_n}\bigr] +\mathcal
    O(D^{\,n_{\max}+2}).
  \end{split}
  \label{eq:H_Weyl_general}
\end{equation}
The totally symmetric tensors $C^{\alpha}_{a_1\dots a_n}$ follow from the Taylor expansion
of the lattice dispersions $\varepsilon^{\alpha}(\bm k)$ about $\bm{k}_0$.
Weyl ordering
$\mathcal{W}[\cdots]$ ensures Hermiticity and guarantees that commutators (gauge-curvature
terms) enter at the correct derivative order.
For the quadratic part ($n=2$), Weyl symmetrization
is automatic because the tensors $C^\alpha_{jk}$ are symmetric.
Explicit symmetrization becomes
necessary only for higher-order terms ($n>2$), such as the quartic $g$-wave sector.

\subsection{Quadratic terms}
For $n=2$, the symmetrized product reduces to $\tfrac{1}{2}\{D_j,D_k\}$, whose structure already
reveals how spin textures generate scalar, gauge, and spin-orbit couplings.
We begin with this
simpler case and then extend the same reasoning to quartic order.

It is convenient to decompose each covariant derivative into longitudinal and transverse parts,
\begin{equation}
  D_j^\parallel = \partial_j - i A_j^\parallel \frac{\sigma_3}{2}, \qquad T_j =
  -\,i\,\frac{\bm A_j^\perp \!\cdot\! \boldsymbol\sigma^\perp}{2},
\end{equation}
so that $D_j = D_j^\parallel + T_j$.
Substituting into the anticommutator gives
\begin{align}
  \label{eq:quad_expanded}
  \begin{split}
    \frac{1}{2}\{D_j , D_k\}
    = \frac{1}{2}\{D_j^\parallel , D_k^\parallel\}
    + \frac{1}{2}\big( D_j^\parallel T_k + T_k D_j^\parallel
    + D_k^\parallel T_j + T_j D_k^\parallel \big)
    + \frac{1}{2}\{T_j , T_k\}.
  \end{split}
\end{align}
This decomposition isolates the three types of contributions of interest.
In the long-wavelength
regime, total derivatives may be discarded and commutators $[D_a^\parallel,f]$ with smooth
functions are suppressed by $(k_F L_\mathrm{txt})^{-1}$.
One then obtains
\begin{equation}
  \frac{1}{2}\{D_j^\parallel , D_k^\parallel\}
  = \big(\partial_j - i A_j^\parallel \tfrac{\sigma_3}{2}\big)
  \big(\partial_k - i A_k^\parallel \tfrac{\sigma_3}{2}\big),
\end{equation}
which describes the emergent Abelian gauge coupling associated with $A_j^\parallel$.
The mixed
terms become
\begin{equation}
  \begin{split}
    \frac{1}{2}\big( D_j^\parallel T_k + T_k D_j^\parallel
    + D_k^\parallel T_j + T_j D_k^\parallel \big)
    = -\,i\big( \bm A_j^\perp \partial_k + \bm A_k^\perp \partial_j \big)
    \cdot \frac{\boldsymbol\sigma^\perp}{2},
  \end{split}
\end{equation}
and represent the texture-induced pseudospin-orbit coupling generated by the transverse connection.
Finally, the purely transverse contribution reads
\begin{equation}
  \frac{1}{2}\{T_j , T_k\}
  = -\,\frac{1}{4}
  \big(\bm A_j^\perp \cdot \bm A_k^\perp\big)\,\sigma_0
  = -\frac{g_{jk}}{4},
\end{equation}
since the transverse Pauli matrices anticommute.
This term therefore acts as a scalar potential
proportional to the texture metric.

The quadratic Weyl product can therefore be written in the transparent form
\begin{equation}
  \mathcal{W}[D_j D_k]
  =
  \frac12\Big(D_jD_k+D_kD_j\Big)
  \approx
  D_j^\parallel D_k^\parallel
  -\frac{i}{2}\Big[
    (\bm A_j^\perp\!\cdot\!\boldsymbol\sigma^\perp)\,D_k^\parallel
    +(\bm A_k^\perp\!\cdot\!\boldsymbol\sigma^\perp)\,D_j^\parallel
  \Big]
  -\frac14\,g_{jk}\,\sigma_0
  \label{eq:W2_final}
\end{equation}
Equivalently, expanding the mixed term explicitly shows that the pieces proportional to
$A^\parallel_j \sigma_3\, T_k$ and $A^\parallel_k \sigma_3\, T_j$ cancel identically because
$\{\sigma_3,\sigma^\perp_\mu\}=0$, while the remaining terms $\propto \partial_j \bm A_k^\perp$
are gradient-suppressed in the long-wavelength limit.
This reduces Eq.~\eqref{eq:W2_final}
to the compact main-text form.

\subsection{Quartic terms}

We now turn to the fourth-order (quartic) symmetrized product of covariant derivatives, which
appears in the long-wavelength expansion of higher-order dispersions.
Such terms involve fully
symmetrized products of the form
\begin{equation}
  \label{eq:W4_def}
  \mathcal{W}[D_j D_k D_\ell D_m]
  = \frac{1}{4!}\sum_{\pi\in S_4}
  D_{a_{\pi(1)}} D_{a_{\pi(2)}} D_{a_{\pi(3)}} D_{a_{\pi(4)}},
\end{equation}
where $(a_1,a_2,a_3,a_4)=(j,k,\ell,m)$, $S_4$ is the symmetric group of all permutations of
$\{1,2,3,4\}$, and $\pi$ denotes one such permutation.
Eq.~\eqref{eq:W4_def} is the Weyl-symmetrized
covariant analogue of the quartic momentum term $k_j k_k k_\ell k_m$.

As in the quadratic case, we decompose each covariant derivative into longitudinal and transverse
parts, $D_a = D_a^\parallel + T_a$.
Keeping terms up to second order in the transverse connection
gives the exact truncation
\begin{align}
  \mathcal{W}[D_j D_k D_\ell D_m]
  =
  \mathcal{W}[D^\parallel_j D^\parallel_k D^\parallel_\ell D^\parallel_m]
  + \sum_{r\in\{j,k,\ell,m\}}
  \mathcal{W}\!\left[T_r \prod_{s\neq r} D^\parallel_s\right]
  \quad
  + \sum_{r<s}
  \mathcal{W}\!\left[T_r T_s \prod_{t\neq r,s} D^\parallel_t\right]
  + \mathcal O(T^3).
  \label{eq:W4_exact_trunc}
\end{align}
Here and below, products such as $\prod_{s\neq r}D^\parallel_s$ are understood in the natural
order inherited from the ordered tuple $(j,k,\ell,m)$; for example,
\begin{equation}
  \prod_{s\neq j} D^\parallel_s = D^\parallel_k D^\parallel_\ell D^\parallel_m,
  \qquad
  \prod_{t\neq j,k} D^\parallel_t = D^\parallel_\ell D^\parallel_m.
  \label{eq:ordered_product_examples}
\end{equation}

Its expansion follows the same formal hierarchy as in the quadratic case, but the physical
distinction between ``electromagnetic'', ``spin-orbit'', and ``scalar'' contributions is no
longer clear-cut.
At this order, longitudinal derivatives appear in higher combinations and
the texture-induced corrections intertwine with the kinetic tensor, so the decomposition is
best regarded as an organizational scheme in powers of the transverse connection rather than
a direct physical separation.

To simplify Eq.~\eqref{eq:W4_exact_trunc}, we use the same long-wavelength assumptions as in the
quadratic case: total derivatives are discarded, and the commutators $[D^\parallel_a,D^\parallel_b]$
and $[D^\parallel_a,T_b]$ are suppressed by $(k_F L_\mathrm{txt})^{-1}\ll 1$.
Under these
assumptions, the zeroth-order piece becomes
\begin{equation}
  \mathcal{W}[D^\parallel_j D^\parallel_k D^\parallel_\ell D^\parallel_m]
  \approx
  D^\parallel_j D^\parallel_k D^\parallel_\ell D^\parallel_m.
\end{equation}

For the terms linear in $T$, consider for example
\begin{equation}
  \mathcal{W}[T_j D^\parallel_k D^\parallel_\ell D^\parallel_m]
  =
  \frac{1}{24}\sum_{\pi\in S_4}
  X_{\pi(1)}X_{\pi(2)}X_{\pi(3)}X_{\pi(4)},
  \qquad
  (X_1,X_2,X_3,X_4)=(T_j,D^\parallel_k,D^\parallel_\ell,D^\parallel_m).
\end{equation}
Grouping the $24$ permutations according to the position occupied by $T_j$, and using that
the six permutations of the three longitudinal factors are equivalent up to the suppressed
commutators, one finds
\begin{align}
  \mathcal{W}[T_j D^\parallel_k D^\parallel_\ell D^\parallel_m]
  \approx
  \frac14\Big(
    T_j D^\parallel_k D^\parallel_\ell D^\parallel_m
    + D^\parallel_k T_j D^\parallel_\ell D^\parallel_m
    + D^\parallel_k D^\parallel_\ell T_j D^\parallel_m
    + D^\parallel_k D^\parallel_\ell D^\parallel_m T_j
  \Big)
  \approx
  T_j D^\parallel_k D^\parallel_\ell D^\parallel_m .
  \label{eq:W4_linear_example}
\end{align}
The same reasoning applies to the other choices of the transverse insertion.
Hence
\begin{equation}
  \sum_{r\in\{j,k,\ell,m\}}
  \mathcal{W}\!\left[T_r \prod_{s\neq r}D^\parallel_s\right]
  \approx
  \sum_{r\in\{j,k,\ell,m\}}
  T_r \prod_{s\neq r}D^\parallel_s .
  \label{eq:W4_linear_final}
\end{equation}

For the terms quadratic in $T$, consider similarly
\begin{equation}
  \mathcal{W}[T_j T_k D^\parallel_\ell D^\parallel_m].
\end{equation}
After moving the longitudinal factors through each other and through the smooth transverse
factors to leading order, the only remaining distinction is the relative ordering of $T_j$
and $T_k$.
Among the $24$ permutations, exactly $12$ place $T_j$ to the left of $T_k$, and
$12$ place $T_k$ to the left of $T_j$.
Therefore
\begin{align}
  \mathcal{W}[T_j T_k D^\parallel_\ell D^\parallel_m]
  \approx
  \frac{12}{24}\,T_j T_k D^\parallel_\ell D^\parallel_m
  +
  \frac{12}{24}\,T_k T_j D^\parallel_\ell D^\parallel_m
  =
  \frac12 \{T_j,T_k\} D^\parallel_\ell D^\parallel_m.
  \label{eq:W4_quadratic_example}
\end{align}
Summing over all unordered pairs gives
\begin{equation}
  \sum_{r<s}
  \mathcal{W}\!\left[T_r T_s \prod_{t\neq r,s}D^\parallel_t\right]
  \approx
  \frac12\sum_{r<s}
  \{T_r,T_s\}\prod_{t\neq r,s}D^\parallel_t .
  \label{eq:W4_quadratic_final}
\end{equation}

Collecting Eqs.~\eqref{eq:W4_exact_trunc}--\eqref{eq:W4_quadratic_final}, and substituting the
definitions of the longitudinal and transverse components, the quartic Weyl product takes the form
\begin{equation}
  \mathcal{W}[D_j D_k D_\ell D_m]
  \approx
  D_j^\parallel D_k^\parallel D_\ell^\parallel D_m^\parallel
  -
  \frac{i}{2}\sum_{r}
  (\bm A_r^\perp\cdot\boldsymbol\sigma^\perp)
  \prod_{s\neq r} D_s^\parallel
  -
  \frac14\sum_{r<s}
  g_{rs}\sigma_0\prod_{t\neq r,s} D_t^\parallel
  +
  \mathcal O\!\big((\bm A^\perp)^3\big).
  \label{eq:W4_final}
\end{equation}

The truncation to second order in $\bm A^\perp$ is justified exactly as in the quadratic case.
If
the texture varies smoothly on a characteristic length scale $L_\mathrm{txt}$, then $|\partial_j\bm
n|\sim 1/L_\mathrm{txt}$ and the expansion parameter is $1/(k_F L_\mathrm{txt})\ll 1$. Since each
additional power of $T_a\sim \partial_a\bm n$ brings an extra factor of $1/(k_F L_\mathrm{txt})$,
the terms linear in $\bm A^\perp$ provide the leading texture-induced correction, the quadratic
metric terms are subleading, and cubic and higher-order contributions become relevant only for
rapidly varying textures.
Eq.~\eqref{eq:W4_final} is the quartic counterpart of Eq.~(8) in the
main text: the same three basic structures reappear, now accompanied by the larger number of
longitudinal derivatives appropriate to fourth order.
In contrast to the quadratic case, the
terms linear in the transverse connection do not admit a further compact simplification after
expanding $D_a^\parallel=\partial_a-iA_a^\parallel\sigma_3/2$, since the different Weyl orderings
generate distinct sign structures rather than cancelling pairwise.

\subsection{$d$-wave and $g$-wave models}
The symmetry of the sublattice matrices $\eta_\alpha$ constrains the derivative tensors
$C^{\alpha}_{j_1\cdots j_n}$ in the inversion-symmetric long-wavelength expansion
\begin{equation}
  H_{\text{kin}} = \sum_{n=2,4,\dots}\frac{(-i)^n}{n!}\, C^{\alpha}_{j_1\cdots j_n}\,
  \eta_\alpha\,\mathcal{W}[D_{j_1}\cdots D_{j_n}],
\end{equation}
where Weyl symmetrization $\mathcal W[\cdots]$ ensures Hermiticity and fixes operator ordering.
The leading nontrivial ranks are $n=2$ (quadratic, ``$d$-wave'') and $n=4$ (quartic, ``$g$-wave'').

The channels $\eta_0$ and $\eta_x$ are symmetry-even on the lattices we consider and therefore
generate only isotropic tensors,
\begin{equation}
  C^{0}_{j_1\cdots j_n}\propto \delta_{(j_1j_2}\cdots\delta_{j_{n-1}j_n)},\qquad
  C^{x}_{j_1\cdots j_n}=K^{x}\,\delta_{(j_1j_2}\cdots\delta_{j_{n-1}j_n)},
\end{equation}
so $\eta_0$ and $\eta_x$ control only the isotropic part of the low-energy expansion.

Nontrivial angular structure arises from $\eta_z$, which is odd under sublattice interchange
and thus permits only selected derivative contractions.
On a tetragonal lattice, the leading
allowed $\eta_z$ term in the $d$-wave case is quadratic and off-diagonal,
\begin{equation}
  C^{z}_{xy}=C^{z}_{yx}=K^{z}, \qquad C^{z}_{xx}=C^{z}_{yy}=0,
\end{equation}
which reproduces the continuum Hamiltonian used in the main text.

In the $g$-wave case, symmetry forbids a quadratic $\eta_z$ term, so the first nonvanishing
contribution occurs at rank $4$.
A minimal tetragonal choice is
\begin{equation}
  C^{z}_{xxyy}=C^{z}_{yyxx}=-C^{z}_{xyxy}=-C^{z}_{yxyx}=K^{z},
\end{equation}
with all other components zero, yielding
\begin{equation}
  C^{z}_{jk\ell m}\,\mathcal{W}[D_j D_k D_\ell D_m] =K^{z}\,\mathcal{W}\!\left[D_x D_y
  (D_x^2-D_y^2)\right],
\end{equation}
which reduces to the $g$-wave form factor $k_x k_y (k_x^2-k_y^2)$ in the texture-free limit.

Retaining terms up to fourth order, the minimal long-wavelength Hamiltonian is
\begin{equation}
  \begin{split}
    H_{g} = C^{x}\eta_x -\frac{1}{2m}\,\eta_0 D^{2} -\frac{K^{x}}{2}\,\eta_x D^{2}
    +\frac{K^{z}}{4!}\, \eta_z\,\mathcal{W}\!\left[D_xD_y(D_x^2-D_y^2)\right] -J\,\sigma_3\eta_z .
  \end{split}
\end{equation}

\section{Low-energy effective theory}
\label{app:low_energy_effective_theory}

We consider the strong-coupling regime where the exchange scale $J$ defines the dominant splitting,
while the sublattice hybridization $C^{x}$ is treated non-perturbatively together with $J$
in the unperturbed Hamiltonian:
\begin{equation}
  H = H_0 + V, \qquad H_0 = C^{x}\eta_x-J\sigma_3\eta_z, \qquad V = H_{\mathrm{kin}} - C^x \eta_x .
\end{equation}
Since $\{\eta_x,\eta_z\}=0$ and $\sigma_3^2=\eta_x^2=\eta_z^2=1$, one has
\begin{equation}
  H_0^2 = \big[(C^x)^2+J^2\big]\,\mathbb{1} \equiv E_0^2\,\mathbb{1}, \qquad E_0=\sqrt{J^2+(C^x)^2}.
\end{equation}
The low- and high-energy projectors (onto energies $\mp E_0$) therefore take the closed form
\begin{equation}
  P=\frac{1}{2}\left(1-\frac{H_0}{E_0}\right),\qquad
  Q=1-P=\frac{1}{2}\left(1+\frac{H_0}{E_0}\right).
\end{equation}

The projection is controlled by this doublet separation rather than by the full electronic
bandwidth.
The isotropic dispersion is proportional to the identity in spin and sublattice space, so
it commutes with $H_0$ and does not mix the two doublets.
The small parameters are instead the spin-
and sublattice-dependent kinetic scales compared with $2E_0$, for example $K^{x,z}k_F^2/(2E_0)\ll1$,
together with the smooth-texture adiabatic condition $v_Fq\ll2E_0$. The Fermi energy, measured from
the low-energy band bottom, must also remain below $2E_0$ so that the remote doublet is unoccupied
in the optical window considered here.

It is convenient to parameterize the sublattice mixing by an angle $\vartheta$,
\begin{equation}
  \cos\vartheta=\frac{J}{E_0},\qquad \sin\vartheta=\frac{C^x}{E_0}.
\end{equation}

For $C^x=0$, the subspace selected by $P$ is spanned by $\{|A\uparrow\rangle,|B\downarrow\rangle\}$.
We fix the projected basis as $\{|A\uparrow\rangle,-|B\downarrow\rangle\}$, which gives the
Pauli matrices
\begin{equation}
  \label{eq:tau_definitions}
  \begin{split}
    \tau_0 & \equiv P\bigl(\eta_0 \sigma_0\bigr)P , \qquad
    \tau_x  \equiv -P\bigl(\eta_x \sigma_x\bigr)P ,\\
    \tau_y & \equiv -P\bigl(\eta_x \sigma_y\bigr)P , \qquad
    \tau_z  \equiv P\bigl(\eta_0 \sigma_3\bigr)P ,
  \end{split}
\end{equation}
The additional useful projection identities are
\begin{equation}
  P\sigma_xP=\sin\vartheta\,\tau_x,\qquad P\sigma_yP=\sin\vartheta\,\tau_y,\qquad
  P\eta_xP=-\sin\vartheta\,\tau_0,\qquad P\eta_zP=\cos\vartheta\,\tau_z,\qquad
  P(\eta_z\sigma_\perp)P=0,
\end{equation}
and
\begin{equation}
  P(\eta_y\sigma_x)P=-\cos\vartheta\,\tau_y,\qquad P(\eta_y\sigma_y)P=\cos\vartheta\,\tau_x.
\end{equation}
Note that several combinations that were nonzero in the full Hilbert space vanish after projection,
e.g.\ $P\eta_yP=0$ and $P(\eta_z\sigma_\perp)P=0$.

To proceed, we decompose the kinetic perturbation into its three channels and further separate
each into spin-flipping and spin-preserving components:
\begin{equation}
  \label{eq:Vsplit}
  V = V_{0,f} + V_{0,p} + V_{z,f} + V_{z,p} + V_{x,f} + V_{x,p}.
\end{equation}

\begin{figure}[hbt]
  \centering
  \includegraphics[]{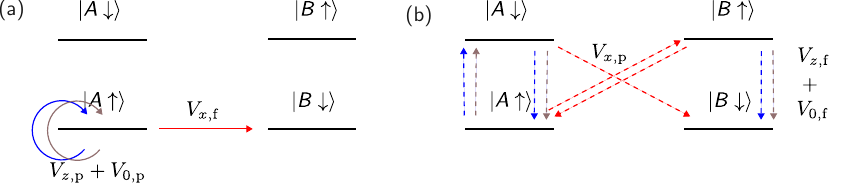}
  \caption{
    Perturbative channels in the comoving-frame doublet.
    (a) First order: $V_{0,p}$ and $V_{z,p}$
    act within a given spin sector, while $V_{x,f}$ flips the physical spin and couples the
    two doublets.
    (b) Second order: only channels with an even number of transverse flips survive
    upon projection, so the leading correction comes from the $\sigma_\perp$--$\sigma_\perp$
    processes built from $V_{0,f}$ and $V_{z,f}$, whereas diagonal $V_{x,p}$ terms are already
    contained in $P H P$.
  }
  \label{fig:perturbations}
\end{figure}

At first order, only perturbations acting inside the low-energy doublet survive, i.e.\ $PVP$.
In particular, in addition to the conventional spin-preserving terms ($V_{0,p}$ and $V_{z,p}$),
there are two texture-induced contributions inside $P$: (i) the rank-2 $\eta_x$ channel generates
the universal pseudospin-orbit coupling term ($V_{x,f}$), and (ii) the isotropic $\eta_0$ channel
contributes an additional pseudospin-orbit coupling term because $P\sigma_\perp P\neq 0$ once
$C^x\neq 0$.

The order of the neglected terms is fixed by the Schrieffer--Wolff expansion.
Since the two
doublets are separated by $2E_0$, the effective Hamiltonian through second order is
\begin{equation}
  H_\mathrm{eff}=-E_0P+PVP-\frac{1}{2E_0}PVQVP
  +\mathcal O\!\left(\frac{\lVert V\rVert^3}{E_0^2}\right).
  \label{eq:Heff_second_order}
\end{equation}
For the hierarchy sweep of Fig.~\ref{fig:multiband_robustness}, $s=K^xk_F^2/(2E_0)$ measures
the inter-doublet mixing.
Truncation after $PVP$ therefore produces a relative $\mathcal O(s)$
error, while retaining the full second-order term removes that leading truncation order and
leaves relative $\mathcal O(s^2)$ corrections to the low-energy Hamiltonian.

Neglecting the constant term $PH_0P=-E_0\,\tau_0$, the projected first-order Hamiltonian can
be written compactly as
\begin{equation}
  \begin{split}
    H_{\mathrm{eff}}^{(1)} = P V P \approx + & \frac{\big(C^0_{j
    k}-\sin\vartheta\,C^x_{jk}\big)\tau_0 + \big(\cos\vartheta\,C^z_{j k}\big)\tau_z}{2}
    \left[ \left(- i \partial_j  - A^\parallel_j \frac{\tau_z}{2}\right) \left(- i
    \partial_k  - A^\parallel_k \frac{\tau_z}{2}\right) + \frac{g_{j k}}{4}\,\tau_0 \right]\\
    - &\frac{\big(C^x_{j k}-\sin\vartheta\,C^0_{jk}\big)}{2}
    \left[
      i \left(\bm A^\perp_j \partial_k
      +\bm A^\perp_k \partial_j \right) \cdot \frac{\bm{\tau}}{2}
    \right].
  \end{split}
\end{equation}

Applying the construction just defined to derive the low-energy Hamiltonian for a spin-textured
$d$-wave altermagnet returns
\begin{equation}
  \begin{split}
    H_{d,\text{eff}} \approx&\; \left(\frac{1}{2m}-\frac{K^x}{2}\sin\vartheta\right)
    \Bigg[ \big(-i\partial_x - A_x^\parallel \tfrac{\tau_z}{2}\big)^2 + \big(-i\partial_y
      - A_y^\parallel \tfrac{\tau_z}{2}\big)^2 + \frac{(\bm A_x^\perp)^2 + (\bm
    A_y^\perp)^2}{4}\,\tau_0 \Bigg] \\[4pt]
    &\;
    -\; i\left(K^x-\frac{\sin\vartheta}{m}\right)
    \big(\bm A_x^\perp \partial_x + \bm A_y^\perp \partial_y\big)\cdot \frac{\boldsymbol\tau}{2}
    \;+\; \frac{K^z \cos\vartheta }{2}\,
    \tau_z
    \left[ -D^\parallel_x D^\parallel_y + \frac{ g_{xy}}{4} \tau_0 \right]\;+\;
    \mathcal{O} \big((\bm{A}_\perp)^3\big).
  \end{split}
  \label{eq:Hd_eff_app}
\end{equation}
which reduces to the expression in the main text for $C^x=0$.

For the $g$-wave, we do the same, but we start from the rank-4 operator
\begin{equation}
  +\frac{K^z}{4!}\,\eta_z\,\mathcal W \big[D_xD_y(D_x^2 - D_y^2)\big].
\end{equation}
We now insert the quartic decomposition and only keep the pieces that survive the projection
$P$.
As in the $d$-wave case, all terms linear in the transverse gauge field carry a factor
$\eta_z\sigma_\perp$, and therefore vanish under $P$ because $P(\eta_z\sigma_\perp)P=0$.
What
does survive are (i) the purely longitudinal quartic piece and (ii) the scalar/metric pieces
$\propto g_{rs}\sigma_0$, because $P(\eta_z\sigma_0)P = \cos\vartheta\,\tau_z$.
Therefore we
obtain, after projection,
\begin{equation}
  \begin{split}
    &P\,\eta_z\,\mathcal{W}[D_xD_y(D_x^2 - D_y^2)]\,P \approx \cos\vartheta\,\tau_z
    \Big\{ D_x^\parallel D_y^\parallel \big[(D_x^\parallel)^2 - (D_y^\parallel)^2\big]
      \label{eq:gwave_core} \\
      &\qquad
      -\frac{3}{4}
      \Big[
        g_{xy}\big((D_x^\parallel)^2 - (D_y^\parallel)^2\big)
        + (g_{xx} - g_{yy})\,D_x^\parallel D_y^\parallel
      \Big]
    \Big\}.
  \end{split}
\end{equation}

Putting this together with the same isotropic quadratic piece and the rank-2 texture-induced
pseudospin-orbit coupling, the projected $g$-wave effective Hamiltonian reads
\begin{equation}
  \begin{split}
    H_{g,\text{eff}} \approx&\; \left(\frac{1}{2m}-\frac{K^x}{2}\sin\vartheta\right)
    \Bigg[ \big(-i\partial_x - A_x^\parallel \tfrac{\tau_z}{2}\big)^2 + \big(-i\partial_y
      - A_y^\parallel \tfrac{\tau_z}{2}\big)^2 + \frac{(\bm A_x^\perp)^2 + (\bm
    A_y^\perp)^2}{4}\,\tau_0 \Bigg] \\[4pt]
    &\;
    -\; i\left(K^x-\frac{\sin\vartheta}{m}\right)
    \big(\bm A_x^\perp \partial_x + \bm A_y^\perp \partial_y\big)\cdot \frac{\boldsymbol\tau}{2}
    \\[4pt]
    &\;
    +\; \frac{K^z \cos\vartheta }{4!}\,
    \tau_z
    \Bigg\{
      D_x^\parallel D_y^\parallel \big[(D_x^\parallel)^2 - (D_y^\parallel)^2\big]
      -\frac{3}{4} \Big[ g_{xy}\big((D_x^\parallel)^2 - (D_y^\parallel)^2\big) + (g_{xx} -
    g_{yy})\,D_x^\parallel D_y^\parallel \Big] \Bigg\} \\[4pt]
    &\;+\; \mathcal{O} \big((\bm{A}_\perp)^3\big).
  \end{split}
  \label{eq:Hg_eff}
\end{equation}

The structure of Eq.~\eqref{eq:Hg_eff} mirrors closely that of the $d$-wave model.
The first
line describes the isotropic $k^{2}$ dispersion dressed by the texture metric, while the second
line is the universal texture-induced pseudospin-orbit coupling term generated by the rank-2
sector, with an additional contribution from the isotropic channel controlled by $\sin\vartheta$.
The last line encodes the characteristic $g$-wave anisotropy with an overall reduction factor
$\cos\vartheta=J/E_0$ arising from the projected $\eta_z$ response.

The accuracy of this projection is tested directly in the numerical robustness checks of
Sec.~\ref{app:numerical_robustness}, where the same optical observables are recomputed in an
unprojected four-band model.

\section{Conductivity in the band basis}
\label{app:kubo_band_basis}
In this section, we recall some principles of the calculation of the optical conductivity
that can be helpful for the understanding of the main results. To make clear which
velocity matrix elements control the dc and finite-frequency responses, it is convenient
to resolve the conductivity bubble in the band basis. Throughout this section, we use
\begin{equation}
  \int_{\bm k}\equiv \int \frac{d^d k}{(2\pi)^d},
  \qquad
  \int_{\varepsilon}\equiv \int \frac{d\varepsilon}{2\pi}.
\end{equation}
Within the bubble approximation, and modeling disorder by a momentum-independent self-energy
$\Sigma^{R/A}=\mp i\Gamma$, the dissipative part of the optical conductivity tensor is
\begin{equation}
  \mathrm{Re}\,\sigma_{ij}(\omega)
  =
  \frac{e^2}{2\omega}
  \int_{\varepsilon,\bm k}
  \Big[f(\varepsilon)-f(\varepsilon+\omega)\Big]\,
  \mathrm{Tr}\!\left[
    v_i(\bm k)\,A(\varepsilon+\omega,\bm k)\,
    v_j(\bm k)\,A(\varepsilon,\bm k)
  \right],
  \qquad
  v_i(\bm k)=\partial_{k_i}H(\bm k),
  \label{eq:sigma_bubble_general}
\end{equation}
valid for $\omega>0$, where $f(\varepsilon)$ is the Fermi function and
\begin{equation}
  A(\varepsilon,\bm k)\equiv i\!\left[G^R(\varepsilon,\bm k)-G^A(\varepsilon,\bm
  k)\right] \label{eq:spectral_function_def}
\end{equation}
is the spectral function.
Introducing the band projectors $P_s(\bm k)=|s,\bm k\rangle\langle
s,\bm k|$, the Green's functions admit the spectral decomposition
\begin{equation}
  G^{R/A}(\varepsilon,\bm k) = \sum_{s=\pm} \frac{P_s(\bm k)}{\varepsilon-E_s(\bm k)\pm
  i\Gamma}, \label{eq:G_band_decomp}
\end{equation}
so that
\begin{equation}
  A(\varepsilon,\bm k) = \sum_{s=\pm} A_s(\varepsilon,\bm k)\,P_s(\bm k), \qquad
  A_s(\varepsilon,\bm k) = \frac{2\Gamma}{\bigl(\varepsilon-E_s(\bm k)\bigr)^2+\Gamma^2}
  = 2\pi\,\delta_\Gamma\!\bigl(\varepsilon-E_s(\bm k)\bigr), \label{eq:A_band_decomp}
\end{equation}
with
\begin{equation}
  \delta_\Gamma(x)\equiv \frac{1}{\pi}\frac{\Gamma}{x^2+\Gamma^2}.
\end{equation}
Likewise, the velocity operator decomposes as
\begin{equation}
  v_i(\bm k) = \sum_{s,s'=\pm} v_i^{ss'}(\bm k)\, |s,\bm k\rangle\langle s',\bm k|,
  \qquad v_i^{ss'}(\bm k)\equiv \langle s,\bm k|\,\partial_{k_i}H(\bm k)\,|s',\bm
  k\rangle . \label{eq:velocity_band_decomp}
\end{equation}
Substituting Eqs.~\eqref{eq:A_band_decomp} and \eqref{eq:velocity_band_decomp} into
Eq.~\eqref{eq:sigma_bubble_general} gives
\begin{equation}
  \mathrm{Re}\,\sigma_{ij}(\omega)
  =
  \frac{e^2}{2\omega}
  \int_{\varepsilon,\bm k}
  \Big[f(\varepsilon)-f(\varepsilon+\omega)\Big]\,
  \sum_{s,s'}
  v_i^{ss'}(\bm k)\,v_j^{s's}(\bm k)\,
  A_s(\varepsilon+\omega,\bm k)\,
  A_{s'}(\varepsilon,\bm k).
  \label{eq:sigma_band_split_omega}
\end{equation}
Eq.~\eqref{eq:sigma_band_split_omega} separates naturally into intraband ($s=s'$) and interband
($s\neq s'$) contributions,
\begin{equation}
  \mathrm{Re}\,\sigma_{ij}(\omega)
  =
  \sigma_{ij}^\mathrm{intra}(\omega)
  +
  \sigma_{ij}^\mathrm{inter}(\omega),
\end{equation}
where
\begin{align}
  \sigma_{ij}^\mathrm{intra}(\omega)
  &=
  \frac{e^2}{2\omega}
  \int_{\varepsilon,\bm k}
  \Big[f(\varepsilon)-f(\varepsilon+\omega)\Big]
  \sum_s
  v_i^{ss}(\bm k)\,v_j^{ss}(\bm k)\,
  A_s(\varepsilon+\omega,\bm k)\,A_s(\varepsilon,\bm k),
  \label{eq:sigma_intra_omega}
  \\[3pt]
  \sigma_{ij}^\mathrm{inter}(\omega)
  &=
  \frac{e^2}{2\omega}
  \int_{\varepsilon,\bm k}
  \Big[f(\varepsilon)-f(\varepsilon+\omega)\Big]
  \sum_{s\neq s'}
  v_i^{ss'}(\bm k)\,v_j^{s's}(\bm k)\,
  A_s(\varepsilon+\omega,\bm k)\,A_{s'}(\varepsilon,\bm k).
  \label{eq:sigma_inter_omega}
\end{align}

In the dc limit $\omega\to 0$, $\tfrac{f(\varepsilon)-f(\varepsilon+\omega)}{\omega}
\;\longrightarrow\; -\partial_\varepsilon f(\varepsilon), $ and
Eq.~\eqref{eq:sigma_band_split_omega} becomes
\begin{equation}
  \sigma_{ij}^\mathrm{DC}
  \equiv \lim_{\omega\to 0}\mathrm{Re}\,\sigma_{ij}(\omega)
  =
  \frac{e^2}{2}
  \int_{\varepsilon,\bm k}
  \bigl(-\partial_\varepsilon f(\varepsilon)\bigr)
  \sum_{s,s'}
  v_i^{ss'}(\bm k)\,v_j^{s's}(\bm k)\,
  A_s(\varepsilon,\bm k)\,A_{s'}(\varepsilon,\bm k).
  \label{eq:sigma_dc_band_split}
\end{equation}
For the intraband piece one has, in the weak-disorder limit,
\begin{equation}
  A_s(\varepsilon,\bm k)^2 = \left[ \frac{2\Gamma}{(\varepsilon-E_s)^2+\Gamma^2}
  \right]^2 \;\xrightarrow[\Gamma\to 0]{}\;
  \frac{2\pi}{\Gamma}\,\delta\!\bigl(\varepsilon-E_s(\bm k)\bigr) \label{eq:Drude_scaling}
\end{equation}
in the sense of distributions.
By contrast, for $s\neq s'$ the product $A_sA_{s'}$ is not
Drude-enhanced and remains nonsingular provided the two bands are not nearly degenerate on
the Fermi surface.
Under this standard clean-limit assumption, the leading dc conductivity
is therefore controlled by the intraband vertices,
\begin{equation}
  \sigma_{ij}^\mathrm{DC}
  =
  \frac{e^2}{2\Gamma}
  \sum_s\int_{\bm k}
  \left(-\partial_\varepsilon f\right)_{\varepsilon=E_s(\bm k)}
  v_i^{ss}(\bm k)\,v_j^{ss}(\bm k).
  \label{eq:sigma_dc_intra_dominant}
\end{equation}

At finite frequency, the interband term in Eq.~\eqref{eq:sigma_inter_omega} becomes resonant
when $\omega$ matches the direct band splitting $\Delta(\bm k)\equiv E_+(\bm k)-E_-(\bm k)$.
For a two-band system and $\omega>0$, the leading interband contribution is
\begin{equation}
  \sigma_{ij}^\mathrm{inter}(\omega)
  =
  \frac{\pi e^2}{\omega}
  \int_{\bm k}
  \Big[f\!\left(E_-(\bm k)\right)-f\!\left(E_+(\bm k)\right)\Big]\,
  v_i^{-+}(\bm k)\,v_j^{+-}(\bm k)\,
  \delta_{2\Gamma}\!\bigl(\omega-\Delta(\bm k)\bigr),
  \label{eq:sigma_optical_inter}
\end{equation}
where $\delta_{2\Gamma}(x)\equiv \frac{1}{\pi}\frac{2\Gamma}{x^2+(2\Gamma)^2}$.
Eq.~\eqref{eq:sigma_optical_inter} shows that the finite-frequency response is governed by the
interband matrix elements $v_i^{+-}$, whereas Eq.~\eqref{eq:sigma_dc_intra_dominant} shows that the
dc response is governed by the intraband vertices $v_i^{ss}$.

In the following sections we evaluate these two quantities explicitly for the helix Hamiltonians:
the intraband identity \eqref{eq:vss_general} controls the dc anisotropy, while the interband
matrix elements $v_i^{+-}$ determine the finite-frequency helix-tracking response.

\section{Intraband velocity vertex and dc conductivity}
\label{app:intraband_vertex}

In this section we show explicitly why the dc conductivity is naturally analyzed in the spiral
basis $(\hat{\bm q},\hat{\bm q}_\perp)$ for both the $d$-wave and $g$-wave helix models, and
then report its helix-orientation dependence in Fig.~\ref{fig:conductivity}.
The static response
is evaluated in the same Kubo-bubble framework as Eq.~\eqref{eq:sigma_dc_band_split}, with velocity
vertices $v_i(\bm k)=\partial_{k_i}H(\bm k)$~\cite{Bonbien_2020_Symmetrized}; in the weak-disorder
dc limit, Eq.~\eqref{eq:sigma_dc_intra_dominant} shows that the leading response is governed
by intraband velocity vertices.

Consider again the generic two-band Hamiltonian
\begin{equation}
  H(\bm k)=d_0(\bm k)\tau_0+d_1(\bm k)\tau_1+d_3(\bm k)\tau_3, \qquad d(\bm
  k)\equiv\sqrt{d_1(\bm k)^2+d_3(\bm k)^2}, \label{eq:two_band_H_repeat}
\end{equation}
with eigenenergies
\begin{equation}
  E_s(\bm k)=d_0(\bm k)+s\,d(\bm k),\qquad s=\pm. \label{eq:Es_def}
\end{equation}
The intraband velocity vertex is defined as
\begin{equation}
  v_i^{ss}(\bm k)\equiv \bra{s}\partial_{k_i}H(\bm k)\ket{s}. \label{eq:vss_def}
\end{equation}
Using the Hellmann--Feynman theorem
\begin{equation}
  v_i^{ss}(\bm k)=\partial_{k_i}E_s(\bm k) =\partial_{k_i}d_0(\bm
  k)+s\,\partial_{k_i}d(\bm k). \label{eq:vss_HF}
\end{equation}
Since $d=\sqrt{d_1^2+d_3^2}$, we obtain the compact identity
\begin{equation}
  v_i^{ss}(\bm k)=\partial_{k_i}d_0(\bm k) +s\,\frac{d_1(\bm k)\,\partial_{k_i}d_1(\bm
  k)+d_3(\bm k)\,\partial_{k_i}d_3(\bm k)}{\sqrt{d_1(\bm k)^2+d_3(\bm k)^2}}. \label{eq:vss_general}
\end{equation}
This is the intraband analogue of the interband identity used in the main text to obtain Eq.~(14).

For the coplanar helix, the long-wavelength $d$-wave Hamiltonian (main text Eq.~(11)) gives
the required derivatives:
\begin{equation}
  \partial_{k_x}d_0=\frac{k_x}{m},\quad \partial_{k_y}d_0=\frac{k_y}{m},\qquad
  \partial_{k_x}d_1=\frac{K^x}{2}q_x,\quad \partial_{k_y}d_1=\frac{K^x}{2}q_y,
  \label{eq:dwave_derivs_1}
\end{equation}
and
\begin{equation}
  \partial_{k_x}d_3=\frac{K^z}{2}k_y,\qquad \partial_{k_y}d_3=\frac{K^z}{2}k_x.
  \label{eq:dwave_derivs_3}
\end{equation}
Substituting into the general identity \eqref{eq:vss_general} gives
\begin{equation}
  v_i^{ss}(\bm k)=\frac{k_i}{m} +s\,\frac{d_1(\bm k)\,\frac{K^x}{2}q_i+d_3(\bm
  k)\,\partial_{k_i}d_3(\bm k)}{\sqrt{d_1(\bm k)^2+d_3(\bm k)^2}}. \label{eq:vss_dwave_full}
\end{equation}
The first term is the isotropic kinetic contribution.
The second term contains the texture
dependence.
Crucially, the part originating from $d_1$ has the universal structure
\begin{equation}
  \delta v_{i,1}^{ss}(\bm k)
  \equiv
  s\,\frac{d_1(\bm k)\,\partial_{k_i}d_1(\bm k)}{\sqrt{d_1(\bm k)^2+d_3(\bm k)^2}}
  =
  s\,\left(\frac{K^x}{2}\right)^2
  \frac{(\bm q\cdot\bm k)\,q_i}{\sqrt{d_1(\bm k)^2+d_3(\bm k)^2}}.
  \label{eq:dv_dwave_qstructure}
\end{equation}
Thus, to leading order in gradients, the helix enters the intraband vertex through the vector
$q_i$ and the scalar $(\bm q\cdot\bm k)$.

\subsection{dc conductivity}
\label{app:dc-conductivity}

When inserted into \eqref{eq:sigma_dc_band_split},
the texture-induced correction can be separated into helix-aligned and crystal-locked parts:
\begin{equation}
  \sigma^\mathrm{DC}_{ij}
  =
  \sigma_\mathrm{iso}\,\delta_{ij}
  +\sigma_\mathrm{an}\left(\hat q_i\hat q_j-\frac{\delta_{ij}}{2}\right)
  +\delta\sigma^\mathrm{cryst}_{ij},
  \label{eq:sigmaDC_tensorform}
\end{equation}
where $\sigma_\mathrm{iso}$ and $\sigma_\mathrm{an}$ are scalar functions of the model parameters,
while $\delta\sigma^\mathrm{cryst}_{ij}$ collects the crystal-harmonic and higher-gradient terms not
fixed by $\hat{\bm q}$ alone.
The first two terms in Eq.~\eqref{eq:sigmaDC_tensorform} are diagonal
in the spiral basis $(\hat{\bm q},\hat{\bm q}_\perp)$; the residual cross component
$\sigma_\times=\hat{\bm q}\cdot\boldsymbol{\sigma}\cdot\hat{\bm q}_\perp$ is carried by
$\delta\sigma^\mathrm{cryst}_{ij}$ and remains subleading in the long-wavelength regime.

\begin{figure}[tbp]
  \centering
  \includegraphics[width=3.4in]{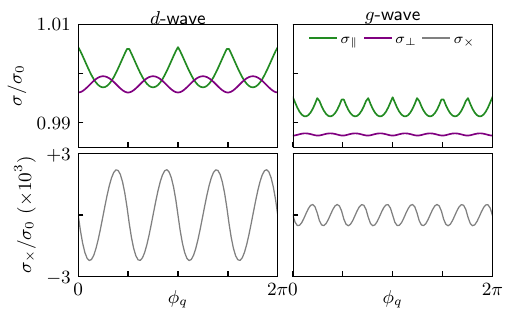}
  \caption{
    Helix-orientation dependence of the dc conductivity in the spiral basis $(\hat{\bm q},\hat{\bm
    q}_\perp)$. Top row: $\sigma_{\parallel}=\hat{\bm q}\cdot\boldsymbol{\sigma}\cdot\hat{\bm
    q}$ and $\sigma_{\perp}=\hat{\bm q}_\perp\cdot\boldsymbol{\sigma}\cdot\hat{\bm
    q}_\perp$, normalized to $\sigma_0=\mu\tau/\pi$. Bottom row: $\sigma_{\times}=\hat{\bm
    q}\cdot\boldsymbol{\sigma}\cdot\hat{\bm q}_\perp$, which captures any residual misalignment
    between $\hat{\bm q}$ and the conductivity principal axes.
    Left: $d$-wave altermagnet, showing
    a pronounced $\sigma_{\parallel}$--$\sigma_{\perp}$ splitting with $\pi/2$-periodic modulation.
    Right: $g$-wave altermagnet, where the $\phi_q$ dependence is a weak $\pi/4$-periodic
    modulation.
    In both cases the cross term $\sigma_{\times}$ remains subleading, so $\hat{\bm q}$
    nearly coincides with the principal axes of the symmetric conductivity tensor.
  }
  \label{fig:conductivity}
\end{figure}

For the $g$-wave helix the two-band form $H=d_0\tau_0+d_1\tau_1+d_3\tau_3$ carries the
\emph{same} $d_0$ and the \emph{same} texture-induced mixing term $d_1=\tfrac{K^x}{2}(\bm
q\!\cdot\!\bm k)$ as the $d$-wave case; only the altermagnetic form factor changes to the
quartic $d_3=\tfrac{K^z}{24}F(\bm k)$ of Eq.~\eqref{eq:Hg_eff}.
Because $d_1$ is unchanged,
its contribution to the intraband vertex retains the universal $\bm q$-aligned structure of
Eq.~\eqref{eq:dv_dwave_qstructure} and feeds the helix-aligned coefficient $\sigma_\mathrm{an}$
in Eq.~\eqref{eq:sigmaDC_tensorform}.
The $g$-wave form factor enters through the weight
$1/\sqrt{d_1^2+d_3^2}$ and the higher-harmonic momentum dependence of $d_3$, whose crystal-locked
contribution is collected in $\delta\sigma^\mathrm{cryst}_{ij}$.
It reweights the angular integrals,
weakening the sensitivity to the helix orientation $\phi_q$ relative to the $d$-wave case, as seen
in Fig.~\ref{fig:conductivity}, without changing the dominance of the $\bm
q$-aligned principal axes.

For a homogeneous N\'eel state ($q=0$), tetragonal symmetry enforces $\sigma_{xx}=\sigma_{yy}$.
A spin helix lowers the symmetry and unlocks an anisotropic correction whose leading long-wavelength
structure is set by $g_{ij}\propto q_i q_j$.
The derivation above therefore explains why it is
natural to analyze the response in the spiral basis defined by $\hat{\bm q}$ and $\hat{\bm
q}_\perp$. As shown in Fig.~\ref{fig:conductivity}, the $d$-wave model exhibits a clear splitting
between $\sigma_{\parallel}$ and $\sigma_{\perp}$ together with a pronounced $\phi_q$ modulation
controlled by the tetragonal anisotropy.
In the $g$-wave model, the conductivity is less sensitive
to the helix orientation, and the dominant effect is an approximately $\phi_q$-independent
anisotropic offset.
In both cases, $\sigma_{\times}$ remains subleading, indicating that $\hat{\bm
q}$ closely tracks the principal axes of the symmetric conductivity tensor for the parameters
shown.
This transport anisotropy is therefore a genuine helix-controlled response, even though
Sec.~\ref{app:afm_discriminator} shows that it is not by itself a qualitative altermagnetic
discriminator.

\section{Interband velocity vertex and helix tracking}
\label{app:vpm_tracking}

In this section we derive the interband velocity matrix elements $v_i^{+-}(\bm k)$ for the
$d$-wave helix Hamiltonian and we show how the helix tracking law follows in the high frequency
regime where the resonant average is approximately $x\leftrightarrow y$ symmetric.

Consider a generic two-band Hamiltonian of the form
\begin{equation}
  H(\bm k)=d_0(\bm k)\,\tau_0+d_1(\bm k)\,\tau_1+d_3(\bm k)\,\tau_3, \qquad d_2(\bm
  k)\equiv 0, \label{eq:two_band_H}
\end{equation}
with band energies $E_\pm(\bm k)=d_0(\bm k)\pm d(\bm k)$ and $d(\bm k)\equiv\sqrt{d_1(\bm
k)^2+d_3(\bm k)^2}$. Introduce the mixing angle $\alpha(\bm k)$ by
\begin{equation}
  \cos\alpha=\frac{d_3}{d},\qquad \sin\alpha=\frac{d_1}{d},
  \qquad \tan\alpha=\frac{d_1}{d_3}.
  \label{eq:alpha_def}
\end{equation}
A convenient choice of normalized eigenvectors of $d_1\tau_1+d_3\tau_3$ is
\begin{equation}
  \ket{+}=
  \begin{pmatrix}
    \cos\frac{\alpha}{2}\\[2pt] \sin\frac{\alpha}{2}
  \end{pmatrix},
  \qquad
  \ket{-}=
  \begin{pmatrix}
    -\sin\frac{\alpha}{2}\\[2pt] \cos\frac{\alpha}{2}
  \end{pmatrix}.
  \label{eq:evecs_choice}
\end{equation}
The velocity operator is $v_i(\bm k)=\partial_{k_i}H(\bm k)$.
Since $\tau_0$ is diagonal in
the band basis, the interband matrix elements depend only on $\partial_{k_i}(d_1\tau_1+d_3\tau_3)$.
Using \eqref{eq:evecs_choice} one finds
\begin{equation}
  \bra{+}\tau_1\ket{-}=\cos\alpha=\frac{d_3}{d},
  \qquad
  \bra{+}\tau_3\ket{-}=-\sin\alpha=-\frac{d_1}{d}.
  \label{eq:tau_offdiag}
\end{equation}
Therefore,
\begin{align}
  v_i^{+-}(\bm k) &\equiv \bra{+}\partial_{k_i}H\ket{-}
  =\partial_{k_i}d_1\,\bra{+}\tau_1\ket{-} +\partial_{k_i}d_3\,\bra{+}\tau_3\ket{-} \nonumber\\
  &=\frac{d_3\,\partial_{k_i}d_1-d_1\,\partial_{k_i}d_3}{\sqrt{d_1^2+d_3^2}}.
  \label{eq:vpm_general}
\end{align}
Eq.~\eqref{eq:vpm_general} is the starting point for both the $d$-wave and $g$-wave helix models.

\subsection{Interband velocity for $d$-wave helix}
\label{app:dwave_vpm}
Using
\begin{equation}
  \partial_{k_x}d_1=\frac{K^x}{2}q_x,\quad
  \partial_{k_y}d_1=\frac{K^x}{2}q_y,\quad
  \partial_{k_x}d_3=\frac{K^z}{2}k_y,\quad
  \partial_{k_y}d_3=\frac{K^z}{2}k_x \,,
\end{equation}
we obtain the exact expressions
\begin{align}
  v_x^{+-}(\bm k) &= \frac{ \frac{K^x}{2}q_x\,d_3 -
    \frac{K^z}{2}k_y\,d_1
  }{
    \sqrt{d_1^2+d_3^2}
  }\,,
  \label{eq:vpm_x_exact}
  \\[4pt]
  v_y^{+-}(\bm k)
  &=
  \frac{
    \frac{K^x}{2}q_y\,d_3
    -
    \frac{K^z}{2}k_x\,d_1
  }{
    \sqrt{d_1^2+d_3^2}
  }\,.
  \label{eq:vpm_y_exact}
\end{align}
Substituting $d_1$ and $d_3$ and keeping terms up to leading order in the helix wave vector,
the numerators simplify due to a cancellation of the $q_x k_x k_y$ and $q_y k_x k_y$ pieces:
\begin{align}
  \frac{K^x}{2}q_x\,d_3-\frac{K^z}{2}k_y\,d_1
  &=
  \frac{K^xK^z}{4}
  \left[
    q_x\!\left(k_xk_y+\frac{q_xq_y}{4}\right)
    -
    k_y(q_xk_x+q_yk_y)
  \right]
  \nonumber\\
  &=
  -\frac{K^xK^z}{4}\,q_y k_y^2
  +
  \frac{K^xK^z}{16}\,q_x^2 q_y\,,
  \label{eq:num_x_expand}
  \\[6pt]
  \frac{K^x}{2}q_y\,d_3-\frac{K^z}{2}k_x\,d_1
  &=
  \frac{K^xK^z}{4}
  \left[
    q_y\!\left(k_xk_y+\frac{q_xq_y}{4}\right)
    -
    k_x(q_xk_x+q_yk_y)
  \right]
  \nonumber\\
  &=
  -\frac{K^xK^z}{4}\,q_x k_x^2
  +
  \frac{K^xK^z}{16}\,q_x q_y^2\,.
  \label{eq:num_y_expand}
\end{align}
In the long-wavelength helix limit $q\ll k_F$, the cubic terms $\sim q^3$ are negligible compared to
the leading $\sim q\,k^2$ terms, so Eqs.~\eqref{eq:vpm_x_exact} and \eqref{eq:vpm_y_exact} reduce to
\begin{align}
  \label{eq:vpm_dwave_final}
  v_x^{+-}(\bm k)
  &\simeq
  -\frac{K^xK^z}{4}\,
  \frac{q_y\,k_y^2}{\sqrt{d_1^2+d_3^2}}\,,
  \qquad
  v_y^{+-}(\bm k)
  \simeq
  -\frac{K^xK^z}{4}\,
  \frac{q_x\,k_x^2}{\sqrt{d_1^2+d_3^2}}\,.
\end{align}
Eqs.~\eqref{eq:vpm_dwave_final} are the $d$-wave long-wavelength result quoted in Eq.~(14) of
the main text and make explicit the key structure $v_x^{+-}\propto q_y$ and $v_y^{+-}\propto q_x$.

\section{Conventional-AFM helix versus altermagnet signatures}
\label{app:afm_discriminator}

We now compare the spin-helix response of an altermagnet (AM) with that of a conventional collinear
antiferromagnet (AFM).
In the helix Hamiltonian, the conventional-AFM limit is obtained by
setting $K^z=0$.
The sublattice-even term
\begin{equation}
  d_1(\bm k)=\frac{K^x}{2}\,\bm q\cdot\bm k
\end{equation}
then remains finite, while the sublattice-odd altermagnetic term $d_3(\bm k)$ vanishes.
Consequently, a conventional AFM helix also has an anisotropic dc response.
The dc anisotropy
alone is therefore not a qualitative altermagnet discriminator.

This concession concerns the existence of anisotropy, not the specific angular response of
the matched minimal model.
With isotropic $d_0(\bm k)$ and $K^z=0$, the helix wave vector $\hat{\bm
q}$ is the only in-plane direction, so rotational covariance sets
$\delta\sigma^\mathrm{cryst}_{ij}=0$ in Eq.~\eqref{eq:sigmaDC_tensorform}.
The tensor form is then
exact and diagonal in the spiral frame, with $\sigma_\times=0$ and an anisotropy magnitude
independent of $\phi_q$: the conventional-AFM response rigidly follows the helix.
The crystal-locked
$d_3\propto K^z k_xk_y$ term breaks this covariance in the altermagnet, producing the fourfold
modulation and finite $\sigma_\times$ of the original calculation.

\begin{figure}[t]
  \centering
  \includegraphics[width=6.8in]{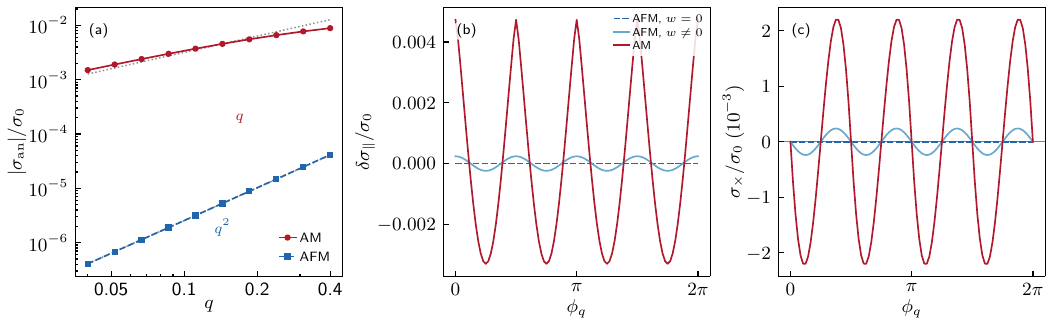}
  \caption{
    Conventional-AFM helix versus altermagnet dc transport.
    (a)~Rotation-invariant dc anisotropy
    $|\sigma_\mathrm{an}|/\sigma_0$ versus helix wave number $q$: the generic long-wavelength
    scaling is $\mathcal{O}(q)$ for the altermagnet (AM) and $\mathcal{O}(q^2)$ for the
    conventional AFM.
    (b,c)~Effect of a sublattice-even $C_4$ band-warping term
    $w(k_x^4+k_y^4)$ ($w=0.02$) on the dc angular structure in the spiral frame: it gives the
    conventional AFM a finite fourfold modulation of the longitudinal $\delta\sigma_\parallel$ (b)
    and a nonzero transverse $\sigma_\times$ (c), partially mimicking the altermagnetic dc
    angular response.
  }
  \label{fig:afm_signatures}
\end{figure}

There is, however, a useful long-wavelength distinction in the magnitude of the dc anisotropy.
To see it, write $\bm q=q\hat{\bm q}$ at fixed helix orientation.
At $K^z=0$, the band splitting is
\begin{equation}
  d(\bm k)=|d_1(\bm k)|=q\,h(\bm k,\hat{\bm q}),\qquad
  \partial_{k_i}d=q\,\partial_{k_i}h, \label{eq:afm_d_qh}
\end{equation}
where $h$ is independent of the pitch $q$.
The two band contributions to the dc conductivity
then take the form
\begin{align}
  \sigma_{ij}(q)\propto\int_{\bm k}
  &A\!\left[\mu-d_0(q^2)-q h\right]
  \left(\partial_i d_0+q\,\partial_i h\right)
  \left(\partial_j d_0+q\,\partial_j h\right)
  \nonumber\\
  &+
  A\!\left[\mu-d_0(q^2)+q h\right]
  \left(\partial_i d_0-q\,\partial_i h\right)
  \left(\partial_j d_0-q\,\partial_j h\right),
  \label{eq:afm_even_integrand}
\end{align}
where $A(x)$ denotes the spectral broadening factor and the explicit $q^2$ dependence of $d_0$
is even in the pitch.
Equivalently,
\begin{equation}
  \sigma_{ij}(q)\propto \int_{\bm k}\left[F_{ij}(\bm k,q)+F_{ij}(\bm k,-q)\right].
  \label{eq:afm_even_q}
\end{equation}
All odd powers of $q$ cancel between the two compensated bands.
Since the $q=0$ tensor is tetragonal
and hence isotropic in the plane, the first anisotropic correction in a conventional AFM helix
is quadratic,
\begin{equation}
  \sigma^\mathrm{an}_\mathrm{AFM}\equiv
  \sqrt{(\sigma_{xx}-\sigma_{yy})^2+4\sigma_{xy}^2}
  =\mathcal{O}(q^2).
  \label{eq:afm_q2}
\end{equation}
When $K^z\neq0$, $d_3$ contains a $q$-independent crystal form factor and this even-in-$q$
cancellation is no longer enforced.
In particular,
\begin{equation}
  \partial_{k_i}d=
  \frac{d_1\,\partial_{k_i}d_1+d_3\,\partial_{k_i}d_3}
  {\sqrt{d_1^2+d_3^2}},
  \label{eq:am_d_derivative}
\end{equation}
contains the interference between the helix-induced $d_1$ channel and the crystal-locked $d_3$
channel.
Near the nodal lines of the $d$-wave form factor, the denominator is controlled by
the helix scale, so this term produces a generic nonanalytic correction linear in the pitch.
For a generic helix direction,
\begin{equation}
  \sigma^\mathrm{an}_\mathrm{AM}=\mathcal{O}(q).
  \label{eq:am_q1}
\end{equation}
The numerical comparison in Fig.~\ref{fig:afm_signatures}(a) confirms these scalings using
the same normalization $\sigma_0=\mu\tau/\pi=\mu/(2\pi\Gamma)$ as in the main text.

The qualitative distinction is sharper in sublattice-odd and interband observables.
The
band-resolved sublattice polarization is controlled by
\begin{equation}
  \langle \tau_3\rangle_\pm=\pm\frac{d_3}{\sqrt{d_1^2+d_3^2}},
  \label{eq:tau3_discriminator}
\end{equation}
and hence vanishes identically at $K^z=0$.
Likewise, the interband velocity matrix element
derived in Eq.~\eqref{eq:vpm_general},
\begin{equation}
  v_i^{+-}(\bm k)= \frac{d_3\,\partial_{k_i}d_1-d_1\,\partial_{k_i}d_3}
  {\sqrt{d_1^2+d_3^2}}, \label{eq:vpm_discriminator}
\end{equation}
also vanishes identically when $d_3=0$.
Since the SOC-free linear dichroism is built from
$v_i^{+-}v_j^{-+}$, the dichroism $\mathfrak{D}(\omega)$ is exactly zero in the conventional-AFM
limit within this nonrelativistic mechanism, in contrast with the finite altermagnetic dichroism
of Fig.~2 of the main text.

To reproduce this angular structure in a conventional AFM, one must add an independent
crystal-locked anisotropy that is absent from the matched isotropic model.
A real tetragonal
band may contain, for example, the sublattice-even crystal warping
\begin{equation}
  d_0(\bm k)\rightarrow d_0(\bm k)+w(k_x^4+k_y^4), \label{eq:tau0_warping}
\end{equation}
which is allowed independently of altermagnetism.
Since $d_0$ enters the intraband velocities,
such a term can generate crystal-locked modulations and a finite $\sigma_\times$ even in the
conventional-AFM limit.
This is shown in Fig.~\ref{fig:afm_signatures}(b,c).
The dc angular
structure is therefore not an absolute qualitative discriminator across arbitrary tetragonal bands.
Nevertheless, the warping preserves the conventional-AFM $\mathcal{O}(q^2)$ scaling, while the
generic altermagnetic contribution is $\mathcal{O}(q)$ and parametrically dominant at small $q$.
Moreover, the same $\tau_0$ warping cannot generate $\tau_3$ polarization or SOC-free dichroism,
because it enters neither Eq.~\eqref{eq:tau3_discriminator} nor Eq.~\eqref{eq:vpm_discriminator}.

\subsection{Symmetry protection of the vanishing interband vertex}
\label{app:afm_symmetry}

The cancellations in Eqs.~\eqref{eq:tau3_discriminator} and \eqref{eq:vpm_discriminator} are
enforced by symmetry.
In spin-group language a conventional collinear antiferromagnet has its two opposite-spin
sublattices connected by a translation, $[C_2\|t_{AB}]$.
In the $d$-wave altermagnet considered here, the corresponding operation contains a proper
rotation, $[C_2\|C_4]$, instead~\cite{Smejkal_2022_Beyond, Smejkal_2022_Emerging}.
The translation requires $[H_\mathrm{kin}(\bm k),\eta_x]=0$, forbidding sublattice-odd kinetic
channels, whereas the rotation permits the $K^zg_n(\bm k)\eta_z$ term of Eq.~(2) of the main text.

For a coplanar texture with plane normal $\bm w$ and no intrinsic SOC, the conventional-AFM
symmetry is $\Sigma=\eta_x\sigma_x$, representing a $\pi$ spin rotation about $\bm w$ composed with
the sublattice-connecting translation.
In the comoving frame, $D_j=k_j-(q_j/2)\sigma_x$ and
$H=\varepsilon(\bm D)\,\eta_0+h(\bm D)\,\eta_x-J\,\sigma_z\eta_z$ with $\varepsilon$ and $h$
arbitrary sublattice-even functions.
Every term commutes with $\Sigma$, as does $v_i=\partial_{k_i}H$, while the two low-energy bands
carry opposite $\Sigma$ eigenvalues; hence $v_i^{+-}\equiv0$.
Any sublattice-even anisotropy, including the warping of Eq.~\eqref{eq:tau0_warping}, preserves this
selection rule, whereas $K^zg_n(\bm D)\eta_z$ anticommutes with $\Sigma$ and breaks it.

\section{Scaling of the crossover frequency}
\label{app:omega_c_scaling}

In this section we derive an analytic estimate for the locked-to-tracking crossover frequency
$\omega_c(\phi_q)$ of the $d$-wave helix and identify the band-structure feature that sets
it.
The result is that all parameter dependence collapses into a single scaling form,
\begin{equation}
  \omega_c \simeq \frac{K^z k_F^{2}}{2}\, X(\rho,\phi_q),
  \qquad
  \rho \equiv \frac{2 K^x q}{K^z k_F},
  \label{eq:omega_c_scaling_form}
\end{equation}
where the dimensionless function $X$ is determined below and $\rho$ measures the competition
between the two gap scales of the problem: the texture-induced $\tau_1$ gap and the altermagnetic
$\tau_3$ gap on the Fermi surface.

\subsection{Reduction to the Fermi circle}

The interband absorption \eqref{eq:sigma_optical_inter} is resonant on the contour $2d(\bm
k)=\omega$, restricted by the occupation factors to a window of width $\sim\omega/2v_F$ around
the Fermi circle.
For frequencies small compared with the Fermi energy the radial integral
can therefore be performed against the resonance, leaving a one-dimensional angular average
over the Fermi circle $\bm k=k_F(\cos\theta,\sin\theta)$,
\begin{equation}
  \mathrm{Re}\,\sigma_{ab}(\omega)\;\propto\;
  \oint \dd\theta\;
  v_a^{+-}(\theta)\,v_b^{-+}(\theta)\;
  \delta_{2\Gamma}\!\big(\omega-2d_{\mathrm{FS}}(\theta)\big),
  \label{eq:sigma_1d}
\end{equation}
with all quantities evaluated at $k=k_F$, where $d_{\mathrm{FS}}(\theta)\equiv
d(k_F\cos\theta,k_F\sin\theta)$ and $\delta_{2\Gamma}$ is the broadened delta function defined
in Eq.~\eqref{eq:sigma_optical_inter}.
On the Fermi circle the two gap components of the helix
Hamiltonian take the form
\begin{equation}
  d_1(\theta)=\Delta_1\cos(\theta-\phi_q), \qquad d_3(\theta)=\Delta_3\sin 2\theta +
  \mathcal O(q^2), \qquad \Delta_1=\tfrac12 K^x q\,k_F, \quad \Delta_3=\tfrac14 K^z
  k_F^{2}, \label{eq:fs_gaps}
\end{equation}
so that the resonance condition involves only the dimensionless gap function
\begin{equation}
  2d_{\mathrm{FS}}(\theta)=2\Delta_3\, g(\theta), \qquad
  g(\theta)=\sqrt{\rho^{2}\cos^{2}(\theta-\phi_q)+\sin^{2}2\theta}, \qquad
  \rho=\frac{\Delta_1}{\Delta_3}=\frac{2K^xq}{K^zk_F}. \label{eq:gap_function}
\end{equation}
Equation~\eqref{eq:sigma_1d} makes the role of frequency explicit: for a given probe frequency,
absorption comes from the points (broadened into arcs by the resonance width $\Gamma$ and by
the thermal occupation window) where the horizontal line $\omega$ intersects the Fermi-surface
splitting $2d_{\mathrm{FS}}(\theta)$.
Thus $\omega$ enters only through the dimensionless level-set
condition $g(\theta)=\omega/(2\Delta_3)$, while the shape of the curve is fixed by $(\rho,\phi_q)$.
This immediately gives the scaling form \eqref{eq:omega_c_scaling_form}.

\subsection{Valley structure and threshold pinning}

\begin{figure}[t]
  \centering
  \includegraphics[width=6.8in]{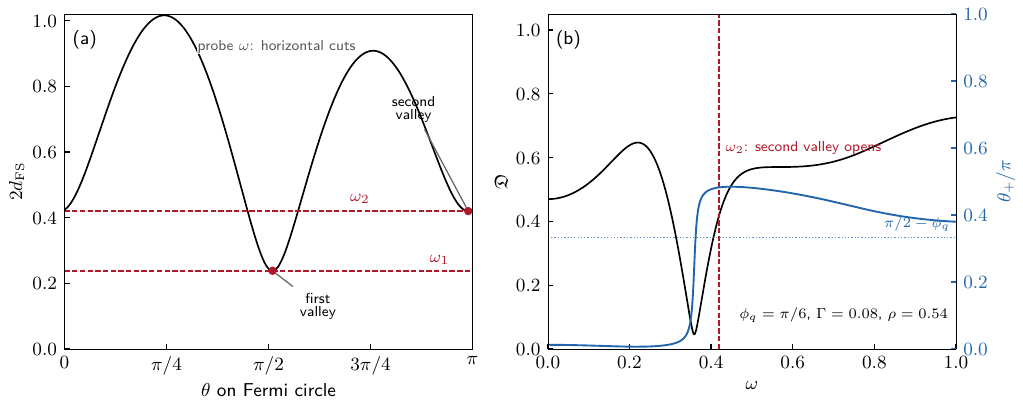}
  \caption{
    Mechanism setting the crossover frequency.
    (a) Fermi-surface splitting
    $2d_{\mathrm{FS}}(\theta)=2\Delta_3 g(\theta)$ of Eq.~\eqref{eq:gap_function} at the band
    parameters of Fig.~2 of the main text ($\phi_q=\pi/6$, $\rho\simeq0.54$), on which it does
    not depend on $\Gamma$.
    The vertical axis is an energy
    scale: a probe at frequency $\omega$ selects the angles where a horizontal line at that height
    intersects the curve.
    The dashed lines mark the two valley openings $\omega_1$ and $\omega_2$.
    (b) Dichroism $\mathfrak D(\omega)$ (black, left axis) and principal axis $\theta_+(\omega)$
    (blue, right axis) from the full two-dimensional Kubo calculation at the same band parameters
    and $\Gamma=0.08$.
    The
    deep minimum of $\mathfrak D$ and the accompanying $\pi/2$ axis rotation occur when the
    second valley starts contributing, near the analytic edge $\omega_2$ [Eq.~\eqref{eq:omega2_def},
    dashed line], with a slight downward shift from finite broadening $\Gamma$.
  }
  \label{fig:omega_c_mechanism}
\end{figure}

For $\rho\lesssim 2$ the gap function $g(\theta)$ possesses two inequivalent local minima per
period: two ``valleys'' produced by the interference of the twofold texture term and the fourfold
altermagnetic term [Fig.~\ref{fig:omega_c_mechanism}(a)].
Absorption at frequency $\omega$ is
carried by the arcs of the level set $g(\theta)=\omega/(2\Delta_3)$.
For $\omega$ below the second
valley opening, this level set samples only the lower valley, so the angular average is dominated by
a small set of anisotropic hotspots near the crystal-axis gap minima and the absorption axis appears
locked.
Once $\omega$ exceeds the second opening, additional resonant arcs enter.
Each newly opened
valley bottom strongly enhances the angular phase-space weight through the factor $1/|g'(\theta)|$,
which has an inverse-square-root threshold dependence in the sharp limit of Eq.~\eqref{eq:sigma_1d};
this new resonant weight competes with the lower-valley channel and drives the dichroism dip.

This structure fixes the crossover mechanism.
Let
\begin{equation}
  \omega_{2}=2\Delta_3\,g_{\min,2}(\rho,\phi_q)
  \label{eq:omega2_def}
\end{equation}
denote the opening of the second valley, with $g_{\min,2}$ its value of $g$ at the corresponding
minimum.
The estimate $\omega_2$ should be understood as the onset scale for the competing
resonant channel, not by itself as a proof of the high-frequency tracking law.
As $\omega$
crosses $\omega_2$, the new threshold-enhanced resonant weight enters with a different polarization
content, the two eigenvalues of the absorptive tensor become nearly degenerate, and the deep
minimum of $\mathfrak D(\omega)$ that defines $\omega_c$ is pinned near this band-structure edge
[Fig.~\ref{fig:omega_c_mechanism}(b)].
At still higher frequencies the resonant arcs sample
the Fermi circle more evenly; then the velocity-vertex argument of Sec.~\ref{app:vpm_tracking},
based on Eqs.~\eqref{eq:vpm_dwave_final}, gives the tracking endpoint.
The crossover frequency
is therefore estimated by the second-valley opening,
\begin{equation}
  \omega_c \simeq \omega_2
  = \frac{K^z k_F^{2}}{2}\; g_{\min,2}(\rho,\phi_q),
  \label{eq:omega_c_estimate}
\end{equation}
where the extrema of $g$ solve $\rho^{2}\sin 2(\theta-\phi_q)=2\sin 4\theta$.
Equation~\eqref{eq:omega_c_estimate} is the explicit form of \eqref{eq:omega_c_scaling_form},
with $X=g_{\min,2}$.

Two limits are worth quoting.
Along the diagonal, $\phi_q=\pi/4$, the extrema equation is solved
by $\sin 2\theta^{*}=-\rho^{2}/4$ and the estimate becomes closed-form,
\begin{equation}
  \omega_c\big|_{\phi_q=\pi/4}
  =\sqrt{2}\,\Delta_1\sqrt{1-\frac{\rho^{2}}{8}}
  =\frac{K^xqk_F}{\sqrt 2}\sqrt{1-\frac{\rho^{2}}{8}}
  \qquad(\rho<2),
  \label{eq:omega_c_pi4}
\end{equation}
while for a weak texture, $\rho\ll1$, at any orientation,
\begin{equation}
  \omega_c\to 2\Delta_1\max\big(|\cos\phi_q|,|\sin\phi_q|\big)
  =K^xqk_F\,\max\big(|\cos\phi_q|,|\sin\phi_q|\big).
  \label{eq:omega_c_smallrho}
\end{equation}
Here the dependence on $\rho$ survives only through $\Delta_1=\rho\,\Delta_3$: for a weak
texture the two valleys sit at the nodes of the altermagnetic form factor ($\sin2\theta=0$,
the crystal axes), where the Fermi-surface gap is purely texture-induced, $g(0)=\rho|\cos\phi_q|$
and $g(\pi/2)=\rho|\sin\phi_q|$, so $\Delta_3$ cancels between the prefactor and $g_{\min,2}$ and
corrections enter at relative order $\rho^{2}$ [cf.~Eq.~\eqref{eq:omega_c_pi4}].
For a finite weak
texture, the crossover scale is therefore linear in the helix wave number.
This scaling cannot be
extrapolated to $q=0$ as a crossover: in the uniform limit the texture-induced dichroism vanishes,
so $\omega_c$ ceases to be defined.

\subsection{Accuracy and range of validity}

\begin{figure}[t]
  \centering
  \includegraphics[width=6.8in]{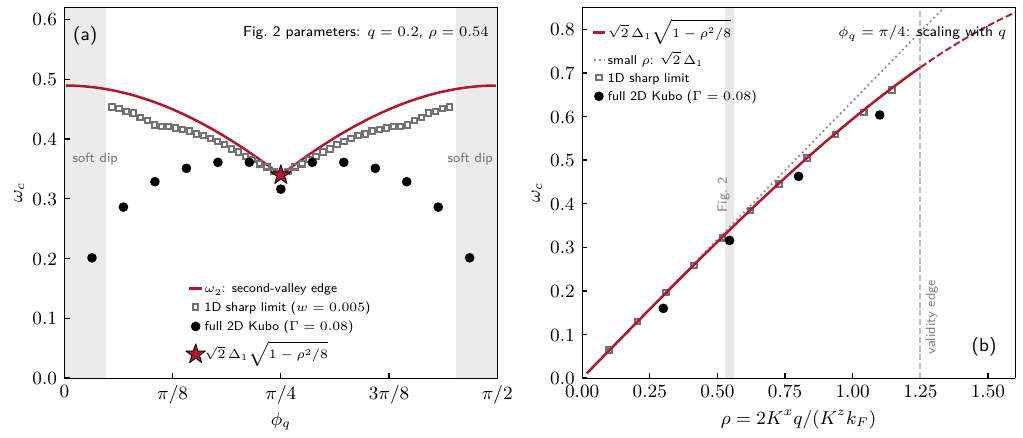}
  \caption{
    Analytic crossover estimate versus numerics.
    (a) $\omega_c(\phi_q)$ at the band parameters
    of Fig.~2 of the main text: the second-valley threshold estimate $\omega_2$ of
    Eq.~\eqref{eq:omega_c_estimate} (red line) and its closed form \eqref{eq:omega_c_pi4} at
    $\phi_q=\pi/4$ (star), compared with the sharp-limit angular average \eqref{eq:sigma_1d}
    (open squares) and the full two-dimensional Kubo calculation at $\Gamma=0.08$
    (filled circles).
    Shaded bands mark the soft-crossover regions near the crystal
    axes, where the dichroism dip is shallow and an $\arg\min_\omega\mathfrak D$ assignment is
    broadening-dominated.
    (b) $\omega_c$ at $\phi_q=\pi/4$ as a function of the helix wave number,
    through $\rho$: the closed form \eqref{eq:omega_c_pi4} (red), its small-$\rho$ linear law
    \eqref{eq:omega_c_smallrho} (dotted), and the two numerical methods.
    The vertical lines
    mark the parameters of Fig.~2 and the validity edge $\rho\simeq1.25$ beyond which the dominant
    avoided crossing migrates to a different feature of $g$.
  }
  \label{fig:omega_c_scaling}
\end{figure}

In the sharp limit $\Gamma\to0$ the estimate \eqref{eq:omega_c_estimate} agrees with the
angular average \eqref{eq:sigma_1d} to better than $1\%$ at $\phi_q=\pi/4$, where the two
valleys open simultaneously and the crossover is sharpest, and at the few-percent level for
intermediate orientations [Fig.~\ref{fig:omega_c_scaling}(a)].
A finite broadening lets resonant
weight from the second valley extend below its sharp threshold, so the observed minimum of
$\mathfrak D$ lies slightly \emph{below} $\omega_2$, with an offset that shrinks as $\sqrt\Gamma$;
at the $\Gamma=0.08$ used here this amounts to a downward shift of order
ten percent [Fig.~\ref{fig:omega_c_scaling}].
The estimate holds for $\rho\lesssim1.25$
[Fig.~\ref{fig:omega_c_scaling}(b)]; for larger $\rho$ the dominant avoided crossing migrates to a
different feature of $g$, and for $\rho>2$ the upper valley disappears altogether.
Exactly along the
crystal axes, $\phi_q=0,\pi/2$, the mirror symmetries force $\sigma_{xy}=0$ and no crossover exists;
approaching the axes the axis rotation completes at $\omega_2$ but becomes increasingly soft,
so an $\arg\min_\omega\mathfrak D$ assignment there is intrinsically broadening-dependent.
The band parameters of Fig.~2 of the main text correspond to $\rho\simeq0.54$.

\subsection{Material-scale estimate}
\label{app:material_estimate}

To translate Eq.~\eqref{eq:omega_c_smallrho} into laboratory units we combine band
parameters of the layered $d$-wave altermagnet \ce{KV2Se2O} with texture length scales
imaged in altermagnets.
A tight-binding fit to first-principles bands, validated against
photoemission~\cite{Jiang_2025_Metallic, Cheng_2026_Realistic}, gives for the
quasi-one-dimensional $d_{xz}/d_{yz}$ doublet a Fermi momentum
$k_F\simeq0.31$~\AA$^{-1}$ and a $d$-wave Fermi-surface splitting
$\Delta_{\mathrm{FS}}\simeq0.5$~eV, where $\Delta_{\mathrm{FS}}$ is the full
separation of the two altermagnetic bands.
Identifying the maximum Fermi-surface splitting as
$\Delta_{\mathrm{FS}}=2\Delta_3=K^zk_F^2/2$, Eq.~\eqref{eq:omega_c_smallrho} can be written as
$\omega_c\simeq(K^x/K^z)\,2\Delta_{\mathrm{FS}}\,(q/k_F)$.
The sublattice-mixing form factor $K^x$ is a spin-preserving inter-sublattice parameter
that spin-locked low-energy fits do not constrain (it connects the low-energy doublet
to its exchange partners) although it is accessible to standard first-principles
calculations; here we assume $K^x\sim K^z$.
N\'eel textures imaged in altermagnetic
\ce{MnTe} have characteristic lengths
$\lambda=2\pi/q\simeq70$--$400$~nm~\cite{Amin_2024_Nanoscale,
Yamamoto_2025_Altermagnetic}, giving $\omega_c\sim5$--$30$~meV ($\approx1$--$7$~THz),
with the locked-to-tracking dichroism spanning the window between $\omega_c$ and the
interband edge $\sim\Delta_{\mathrm{FS}}$, i.e., terahertz to mid-infrared.
This is an
order-of-magnitude guide rather than a material prediction: the Fermi surface of
\ce{KV2Se2O} (open quasi-one-dimensional sheets plus hole pockets at the zone boundary)
differs from the closed Fermi surface of the minimal model, and the bulk magnetic
stacking of this family is still debated.

\section{Intrinsic spin-orbit coupling}
\label{app:soc}

In this section we extend the four-band model to include intrinsic (relativistic) spin-orbit
coupling (SOC) and derive its effect on the low-energy doublet.
The result is that, at first
order in the SOC couplings, intrinsic SOC enters the projected theory \emph{exclusively} as an
additive shift of the two emergent gauge fields of Sec.~\ref{app:su2_gauge_formalism}.
For a planar
helix the shift amounts to a renormalization of the helix wave vector, so the effective theory is
form-invariant and the locked-to-tracking crossover survives.

\subsection{Symmetry-allowed channels}

To linear order in $\bm k$ the two spin-orbit invariants are the Rashba pseudoscalar
$(\bm k\times\bm\sigma)_z=k_x\sigma_y-k_y\sigma_x$ and the scalar $\bm
k\cdot\bm\sigma_\parallel=k_x\sigma_x+k_y\sigma_y$; both are odd under the horizontal mirror
$M_z$, so either requires structural inversion asymmetry.
Requiring invariance under time reversal
$\mathcal T$, the altermagnetic antiunitary $[\mathcal T C_{4z}\eta_x]$, and the vertical mirror
combined with the sublattice swap, $M_x\eta_x$, leaves three channels,
\begin{equation}
  H_\mathrm{SOC} =\lambda_A\,\eta_0\,(\bm k\times\bm\sigma)_z +\lambda_B\,\eta_x\,(\bm
  k\times\bm\sigma)_z +\lambda_C\,\eta_y\,\bigl[\bm k\times(\bm
  n\times\bm\sigma)\bigr]_z, \label{eq:Hsoc_channels}
\end{equation}
with $[\bm k\times(\bm n\times\bm\sigma)]_z=n_z(\bm k\cdot\bm\sigma_\parallel)-(\bm k\cdot\bm
n)\sigma_z$ and all momenta and spin components in the crystal frame. Channels A and B are
the N\'eel-\emph{even} Rashba couplings; symmetry allows both the uniform ($\eta_0$) and the
bond ($\eta_x$) sublattice structure, while the staggered Rashba $\eta_z(\bm k\times\bm\sigma)_z$
is forbidden by the altermagnetic antiunitary.
Channel C is N\'eel-\emph{odd}: since $\eta_y$
is odd under time reversal, this channel must carry an odd power of the N\'eel order, meaning that
$\lambda_C$ is an odd function of the exchange, $\lambda_C\propto J$ at leading order, and
the N\'eel direction enters the operator itself through the covariant triple product, which
reduces to the uniform-N\'eel form $\eta_y(\bm k\cdot\bm\sigma_\parallel)$ for $\bm n=\hat{\bm
z}$. While symmetry alone would permit the two pieces $n_z(\bm k\cdot\bm\sigma_\parallel)$ and
$(\bm k\cdot\bm n)\sigma_z$ with independent coefficients $(a,b)$, only their combination as
$(a-b)/2\equiv \lambda_C$ enters the projected theory at first order, so none of the results
below depend on the triple-product locking.
A microscopically derived model confirms this form
of the SOC Hamiltonian~\cite{AbadilloUriel_2026_All}.

\subsection{Rotation of the spin-orbit vertex to the comoving frame}
\label{app:soc_comoving}

Throughout this section the geometric objects $U(\bm r)$, $R(\bm r)$, the triad $(\bm e_1(\bm
r),\bm e_2(\bm r),\bm n(\bm r))$, and the connection $\bm A_j^\perp(\bm r)$ are position-dependent
through the texture, whereas the crystal axes $\hat{\bm x}_j$ ($j=x,y$) and $\hat{\bm z}$,
the couplings $\lambda_{A,B,C}$, and the sublattice-mixing angle $\vartheta$ ($\cos\vartheta=J/E_0$,
$\sin\vartheta=C^x/E_0$) are constants.

Before projecting, it is instructive to write $H_\mathrm{SOC}$ in the comoving frame set by
the local rotation $U$ of Eq.~\eqref{eq:unitary_definition}.
Since $U$ acts only on spin, it
commutes with the sublattice structure $\eta_a$ and with the crystal-momentum factors before
covariantization, so the conjugation reduces to the adjoint action on the Pauli vector.
This
is precisely the $\mathrm{SO(3)}$ rotation $R_{ij}$ of Eq.~\eqref{eq:rotation_from_U}: from
$U\sigma_j U^\dagger=R_{ij}\sigma_i$ and the orthogonality of $R$, the laboratory Pauli vector
is rigidly rotated onto the comoving frame,
\begin{equation}
  U^\dagger\bm\sigma\,U=R\bm\sigma =\bm e_1\sigma_1+\bm e_2\sigma_2+\bm n\sigma_3 ,
  \label{eq:soc_spin_rotation}
\end{equation}
where the last form uses that the columns of $R$ are the comoving triad $(\bm e_1,\bm e_2,\bm
n)$ in the laboratory frame. Momenta simultaneously become covariant, $\bm k\to\bm\pi$, with
$\bm\pi$ the gauge-covariant kinetic momentum of Sec.~\ref{app:su2_gauge_formalism}.

Since $H_\mathrm{SOC}$ depends on spin only through $\bm\sigma$ and the crystal couplings
$\lambda_{A,B,C}$ are constants unaffected by the rotation, conjugation acts by the single
substitution $\bm\sigma\to R\bm\sigma$ of Eq.~\eqref{eq:soc_spin_rotation}.
Applying it to
the three channels of Eq.~\eqref{eq:Hsoc_channels} gives the compact comoving-frame form
\begin{equation}
  \begin{split}
    U^\dagger H_\mathrm{SOC}\,U =&\bigl(\lambda_A\eta_0+\lambda_B\eta_x\bigr)
    \bigl(\bm\pi\times R\bm\sigma\bigr)_z +\lambda_C\,\eta_y\, \bigl[\bm\pi\times(\bm
    n\times R\bm\sigma)\bigr]_z \\
    =&\bigl(\lambda_A\eta_0+\lambda_B\eta_x\bigr) \Bigl[(\bm\pi\times\bm
      e_1)_z\,\sigma_1+(\bm\pi\times\bm e_2)_z\,\sigma_2 +(\bm\pi\times\bm
    n)_z\,\sigma_3\Bigr]\\ &+\lambda_C\,\eta_y\Bigl[(\bm\pi\times\bm e_2)_z\,\sigma_1
    -(\bm\pi\times\bm e_1)_z\,\sigma_2\Bigr],
  \end{split}
  \label{eq:soc_comoving}
\end{equation}
that is, $H_\mathrm{SOC}$ with its spin vector rotated by $R$.

All texture dependence now resides in the rotation $R$, in the triad, and in the connection
hidden in $\bm k\to\bm\pi$, while the crystal couplings $\lambda_{A,B,C}$ carry no texture
dependence.
Two structural facts, visible already here, control the projection below: (a) channel
C carries no $\sigma_3$ component (the triple product is pointwise transverse to $\bm n$) and in the
comoving frame it is exactly channel B rotated by $\pi/2$ about $\sigma_3$ (with $\eta_x\to\eta_y$),
which makes the pointwise B$\,\equiv\,$C fusion on the doublet manifest; (b) the $\sigma_3$ pieces
survive projection only through the $\eta_0$ channel A (giving a $\tau_z$ term), whereas the
$\eta_x$ one connects to the $J$-split sector and enters only at second order.

\subsection{SOC as a shift of the emergent gauge fields}

We now project the comoving-frame Hamiltonian Eq.~\eqref{eq:soc_comoving} onto the low-energy
doublet.
Applying the projection identities of Sec.~\ref{app:low_energy_effective_theory} gives, at
first order in the SOC couplings and to leading order in gradients, the pointwise identity
\begin{equation}
  P\,U^\dagger H_\mathrm{SOC}\,U\,P =-\bar\alpha(\vartheta)\, \Bigl[(\bm\pi\times\bm
  e_1)_z\,\tau_x+(\bm\pi\times\bm e_2)_z\,\tau_y\Bigr]
  +\bar\lambda_A(\vartheta)\,(\bm\pi\times\bm n)_z\,\tau_z , \label{eq:soc_projected}
\end{equation}
with the $\vartheta$-rotated weights
\begin{equation}
  \bar\alpha(\vartheta)=\lambda_B+\cos\vartheta\,\lambda_C
  -\sin\vartheta\,\lambda_A,
  \qquad
  \bar\lambda_A(\vartheta)=\lambda_A-\sin\vartheta\,\lambda_B .
  \label{eq:soc_weights}
\end{equation}
The A and (B,\,C) channels mix under the sublattice-mixing angle exactly as the $\eta_0$ and
$\eta_x$ kinetic channels do in Eq.~\eqref{eq:Hd_eff_app}; for $C^x=0$ the weights reduce to
$\bar\alpha=\lambda_B+\lambda_C$ and $\bar\lambda_A=\lambda_A$.

Equation~\eqref{eq:soc_projected} states that the projected SOC occupies precisely the two
coupling slots used by the texture itself.
Setting $C^x=0$ for compactness and comparing with
Eq.~\eqref{eq:Hd_eff_app}, it is an additive renormalization of the emergent connections,
\begin{equation}
  \begin{aligned}
    \bm A_j^{\perp}&\to\bm A_j^{\perp}+\delta\!\bm A_j^{\perp},\quad
    \delta\!\bm A_j^{\perp}=-\frac{2\bar\alpha}{K^x}
    \bigl[(\hat{\bm z}\times\hat{\bm x}_j)
    -\bigl(\bm n\cdot(\hat{\bm z}\times\hat{\bm x}_j)\bigr)\,\bm n\bigr],\\
    A_j^{\parallel}&\to A_j^{\parallel}+\delta\!A_j^{\parallel},\quad
    \delta\!A_j^{\parallel}=-2m\bar\lambda_A\,\bigl(\bm n\times\hat{\bm z}\bigr)_j .
  \end{aligned}
  \label{eq:soc_deltaA}
\end{equation}
The structure of Eq.~\eqref{eq:soc_deltaA} makes the SOC--texture interplay sharp.
The texture
connection $\bm A_j^\perp=\bm n\times\partial_j\bm n$ has its direction set by the texture
and magnitude $\mathcal O(\partial\bm n)$; the SOC shift is \emph{crystal-pinned} in direction,
$\hat{\bm z}\times\hat{\bm x}_j$, gradient-independent in magnitude, and acquires texture
sensitivity only through the local transverse projection onto the plane orthogonal to $\bm n$.
Three structural consequences follow directly: (i) channels B and C project to the \emph{same}
operator, so only the combination $\bar\alpha$ is observable at first order; (ii) channel A
shifts the Abelian connection, with gauge-invariant content $\delta b=(\nabla\times\delta\!\bm
A^\parallel)_z =2m\bar\lambda_A\,\nabla\!\cdot\bm n_\parallel$ --- zero on average for any
planar helix; (iii) no $\tau_0$ and no $k$-independent (gap) terms are generated, and for an
in-plane texture ($n_z=0$, $\bm e_1=\hat{\bm z}$) the $\tau_x$ component vanishes pointwise.

\subsection{Renormalization of the helix wave vector}

For the planar helix $\bm n=\bm u\cos\theta+\bm v\sin\theta$, $\theta=\bm q\cdot\bm r$, with
plane normal $\bm w=\bm u\times\bm v$ and the frame choice $\bm e_1=\bm w$ used in the main
text, the comoving triad is $\bm e_1=\bm w$ (constant), $\bm e_2=\bm n\times\bm w$, and $\bm
e_3=\bm n$, so Eq.~\eqref{eq:soc_projected} reads pointwise
\begin{equation}
  P\,U^\dagger H_\mathrm{SOC}\,U\,P =-\bar\alpha\,(\bm\pi\times\bm w)_z\,\tau_x
  -\bar\alpha\,\bigl[\bm\pi\times(\bm n\times\bm w)\bigr]_z\,\tau_y
  +\bar\lambda_A\,(\bm\pi\times\bm n)_z\,\tau_z . \label{eq:soc_helix_pointwise}
\end{equation}

Only the $\tau_x$ term in Eq.~\eqref{eq:soc_helix_pointwise} is independent of the helix phase.
The winding $\tau_y$ and $\tau_z$ terms have zero phase average and contain only the Fourier
harmonics $e^{\pm i\bm q\cdot\bm r}$, which couple $\bm k$ to the sidebands $\bm k\pm\bm q$ but have
no first-order matrix element within a fixed momentum sector.
The first-order, momentum-diagonal
SOC correction is therefore
\begin{equation}
  \bigl\langle P\,U^\dagger H_\mathrm{SOC}\,U\,P\bigr\rangle_\theta
  =
  -\bar\alpha\,(\bm\pi\times\bm w)_z\,\tau_x .
  \label{eq:soc_helix_avg}
\end{equation}

At first order intrinsic SOC is thus absorbed into a renormalized helix wave vector in the
$\tau_x$ channel: using $-(\bm\pi\times\bm w)_z=(\hat{\bm z}\times\bm w)\cdot\bm\pi$, the total
$\tau_x$ coupling is $\tfrac{K^x}{2}(\bm q_\mathrm{eff}\cdot\bm\pi)$ with
\begin{equation}
  \bm q_\mathrm{eff}
  =\bm q+\frac{2\bar\alpha}{K^x}\,\bigl(\hat{\bm z}\times\bm w\bigr).
  \label{eq:soc_qeff}
\end{equation} Note what does \emph{not} shift: the
$\tau_z$ texture offset $q_xq_y/4$ and the scalar $q^2/8m$ retain the bare $\bm q$ (the texture
metric $g_{jk}=q_jq_k$ is SOC-blind at this order), so the two texture scales that coincide
in the SOC-free theory misalign at order $\zeta$, where
\begin{equation}
  \zeta=\frac{2\bar\alpha}{K^xq}\,\bigl|\bm w\times\hat{\bm z}\bigr|
  \label{eq:soc_zeta}
\end{equation}
is the dimensionless ratio of the SOC shift to the texture gauge field.

Intrinsic SOC makes the helix-plane normal $\bm w$ observable.
In the SOC-free theory this
orientation can be changed by a global spin rotation; with SOC, Eq.~\eqref{eq:soc_qeff} couples
it to the crystal axis through $\delta\bm q\equiv\bm q_\mathrm{eff}-\bm q=(2\bar\alpha/K^x)(\hat{\bm
z}\times\bm w)$. The resulting geometry dependence is:
\begin{itemize}
  \item \emph{in-plane spiral} ($\bm w=\pm\hat{\bm z}$; $\bm n$ rotates in the crystal
    plane): $\delta\bm q=0$ --- the effective-$\bm q$ shift vanishes identically, while the
    winding harmonics remain in the full model;
  \item \emph{proper screw} of chirality $s=\pm1$ ($\bm w=s\hat{\bm q}$):
    $\delta\bm q\perp\bm q$ and $\bm q_\mathrm{eff}/q=\hat{\bm q}+s\zeta(\hat{\bm z}\times\hat{\bm
    q})$. Hence the effective helix vector has magnitude $q\sqrt{1+\zeta^2}$ and azimuth
    $\phi_\mathrm{eff}=\phi_q+\arctan(s\zeta)$.
    Because the high-frequency tracking law is
    $\theta_+\simeq\pi/2-\phi_\mathrm{eff}$, the absorption axis shifts by
    $\delta\theta_+=-\arctan(s\zeta)=\mp\arctan\zeta$;
  \item \emph{cycloid} ($\bm w=\pm\,\hat{\bm z}\times\hat{\bm q}$): $\delta\bm
    q\parallel\bm q$ --- pure renormalization of the helix wave number,
    $q_\mathrm{eff}=(1\mp\zeta)\,q$, with the sign set by the helix handedness.
\end{itemize}
The shift $\delta\bm q$ is odd under helix-chirality reversal ($\bm w\to-\bm w$) and even under
N\'eel reversal ($\bm n\to-\bm n$, which leaves $\bm w$ invariant): intrinsic SOC makes the
response sensitive to the texture handedness, to which the SOC-free theory is blind.

The form of the $\tau_x$ channel is preserved under $\bm q\to\bm q_\mathrm{eff}$, so the
crossover survives at first order.
The low-frequency locked regime, set by the crystal ($k_xk_y$)
structure of the $\tau_z$ channel, is untouched; the high-frequency endpoint follows
$\theta_+\simeq\pi/2-\phi_\mathrm{eff}$, with $\phi_\mathrm{eff}=\arg\bm q_\mathrm{eff}$; and,
through the scaling of Sec.~\ref{app:omega_c_scaling} ($\omega_c\propto q$ at small $\rho$),
the crossover frequency is rescaled by the cycloid,
$\omega_c\to(1\mp\zeta)\,\omega_c$, and unchanged at $\mathcal O(\zeta)$ for the
screw and in-plane geometries.
For cycloid chirality $s=\pm1$ (with $s=+$ the compressed
branch) and $\zeta<1$, the dotted trends in Fig.~4 follow directly from the weak-texture scaling
\begin{equation}
  \frac{\omega_c(\zeta,s)}{\omega_c(0)}
  =\frac{|\bm q_\mathrm{eff}|}{q}+\mathcal O(\rho^2\zeta,\zeta^2)
  =|1-s\zeta|+\mathcal O(\rho^2\zeta,\zeta^2)
  =1-s\zeta+\mathcal O(\rho^2\zeta,\zeta^2).
  \label{eq:soc_cycloid_trend}
\end{equation}

The analysis above holds at first order in the SOC couplings and within the projected
two-band regime, $\lambda_\mathrm{SOC}k_F\ll2E_0$, as discussed in
Sec.~\ref{app:low_energy_effective_theory}.
The zero-Fourier component of the projected SOC
produces the form-preserving replacement $\bm q\to\bm q_\mathrm{eff}$, whereas the winding
$\tau_y$ and $\tau_z$ components have zero phase average and couple the momentum sector $\bm k$
to the sidebands $\bm k\pm\bm q$.
Away from sideband degeneracies, their leading momentum-diagonal
contribution is therefore second order in the SOC couplings.
Near sideband crossings, however, they
can produce resonant mixing that is not captured by a simple renormalization of $\bm q$.
Thus the
effective-$\bm q$ description should be understood as the leading weak-SOC reduction rather than an
exact elimination of the winding harmonics.
Finally, the first-order description assumes $\zeta\ll1$,
so the uniform limit $q\to0$ at fixed SOC couplings falls outside the controlled regime.
In that limit a crystal-pinned background can remain in the optical response,
but such a background carries neither the locked-to-tracking crossover
nor the helix-chirality dependence of the texture-generated signatures.

\section{Numerical robustness checks}
\label{app:numerical_robustness}

\begin{figure}[t]
  \centering
  \includegraphics[width=6.8in]{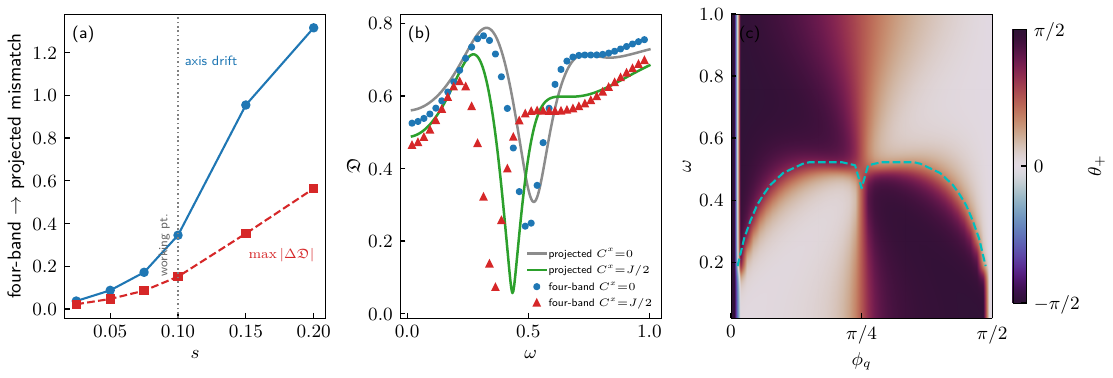}
  \caption{
    Multiband robustness of the projected two-band optics.
    Exact zero-SOC four-band Kubo results (markers; all interband channels included) are compared
    with the projected theory (lines).
    (a)~The principal-axis drift and maximum dichroism mismatch at $\phi_q=\pi/6$ scale linearly
    with the hierarchy parameter $s=K^xk_F^2/(2E_0)$; the working point $s=0.1$ ($\mu=2$) is
    marked.
    (b)~At $C^x=0$, the four-band dichroism follows the bare projected theory, while at
    $C^x=J/2$ it follows the appropriately renormalized projected theory.
    (c)~The exact principal axis reproduces the locked-to-tracking rotation; dashed lines show the
    projected crossover, and mirror orientations are masked.
    At the working point, $\omega_c$ agrees within $\simeq5\%$ and the axis within
    $\lesssim0.4$~rad.
  }
  \label{fig:multiband_robustness}
\end{figure}

The analytic sections above use two controlled reductions of the microscopic helix problem:
projection to the low-energy spin-sublattice doublet, and a weak-SOC treatment in which the
momentum-diagonal effect is a shift of the emergent gauge field while winding terms enter
perturbatively.
We therefore collect the direct numerical checks of these approximations here,
immediately before the parameter summary: Fig.~\ref{fig:multiband_robustness} tests the multiband
stability of the projected optical response, while Fig.~\ref{fig:soc_robustness} tests the weak-SOC
downfolding against a four-band helix calculation with all winding terms retained.
The details of
the method are described in what follows.

\subsection{Superlattice calculation with intrinsic spin-orbit coupling}
\label{sec:sm_soc_superlattice}

To test the robustness of the optical crossover beyond the projected weak-SOC treatment, we
also diagonalize the full four-band helix problem in the laboratory frame.
This avoids expanding
the winding spin-orbit terms in powers of the sideband coupling.
The calculation is performed
for the $d$-wave continuum model
\begin{equation}
  H(\mathbf k,\mathbf r) = H_\mathrm{inv}(\mathbf k) -
  J\,\eta_z\,\mathbf n(\mathbf r)\cdot\boldsymbol\sigma
  +
  H_\mathrm{SOC}(\mathbf k,\mathbf r),
\end{equation}
with
\begin{equation}
  H_\mathrm{inv}(\mathbf k) = \frac{k^2}{2m}\eta_0 + \left(C^x + \frac{K^x}{2}k^2
  \right)\eta_x + \frac{K^z}{2}k_x k_y \eta_z .
\end{equation}
Here $\eta_\alpha$ and $\sigma_i$ act in sublattice and spin space, respectively.
The intrinsic
SOC terms retained in the calculation are
\begin{align}
  H_\mathrm{SOC} =&\; \lambda_A \eta_0 (k_x\sigma_y-k_y\sigma_x) + \lambda_B \eta_x
  (k_x\sigma_y-k_y\sigma_x) \nonumber\\
  &+
  \lambda_C \eta_y
  \left[
    n_z(\mathbf r)(k_x\sigma_x+k_y\sigma_y)
    -
    (\mathbf k\cdot\mathbf n(\mathbf r))\sigma_z
  \right].
  \label{eq:sm_soc_lab}
\end{align}
The $A$ and $B$ channels are independent of the N\'eel texture, whereas the $C$ channel winds
with $\mathbf n(\mathbf r)$.

We specialize to a single-$q$ coplanar helix,
\begin{equation}
  \mathbf n(\mathbf r)
  =
  \mathbf u\cos\theta+\mathbf v\sin\theta,
  \qquad
  \theta=\mathbf q\cdot\mathbf r,
  \qquad
  \mathbf w=\mathbf u\times\mathbf v .
\end{equation}
Its only Fourier components are
\begin{equation}
  \mathbf n(\mathbf r)
  =
  \mathbf a_+ e^{i\theta}
  +
  \mathbf a_- e^{-i\theta},
  \qquad
  \mathbf a_+=\frac{\mathbf u-i\mathbf v}{2},
  \qquad
  \mathbf a_-=\mathbf a_+^* .
\end{equation}
Thus the helix defines a one-dimensional superlattice along $\hat{\mathbf q}$.
For a Bloch
momentum $\mathbf k$ in the reduced Brillouin zone (RBZ), $k_\parallel\in[-q/2,q/2)$, we introduce
plane-wave rungs
\begin{equation}
  \boldsymbol\kappa_n=\mathbf k+n\mathbf q,
  \qquad
  n\in\mathbb Z ,
\end{equation}
and use the basis
\begin{equation}
  |n,\eta,\sigma\rangle = e^{i(\mathbf k+n\mathbf q)\cdot\mathbf r} |\eta,\sigma\rangle .
\end{equation}
In this basis the Hamiltonian is block tridiagonal in the rung index.
The diagonal blocks are
\begin{equation}
  H_{nn} = H_\mathrm{inv}(\boldsymbol\kappa_n) + \lambda_A\eta_0
  (\kappa_{n,x}\sigma_y-\kappa_{n,y}\sigma_x) + \lambda_B\eta_x
  (\kappa_{n,x}\sigma_y-\kappa_{n,y}\sigma_x).
\end{equation}
The exchange texture and the $C$-channel SOC connect only neighboring rungs:
\begin{equation}
  H_{n+1,n} = -J\eta_z\,\mathbf a_+\cdot\boldsymbol\sigma + \lambda_C\eta_y
  S_C(\boldsymbol\kappa_{n+1/2},\mathbf a_+), \label{eq:sm_ladder_hop}
\end{equation}
with
\begin{equation}
  \boldsymbol\kappa_{n+1/2}
  =
  \mathbf k+\left(n+\frac12\right)\mathbf q
\end{equation}
and
\begin{equation}
  S_C(\boldsymbol\kappa,\mathbf a) = a_z(\kappa_x\sigma_x+\kappa_y\sigma_y) -
  (\kappa_x a_x+\kappa_y a_y)\sigma_z .
\end{equation}
The reverse hopping is fixed by Hermiticity,
\begin{equation}
  H_{n,n+1}=H_{n+1,n}^\dagger .
\end{equation}
The midpoint momentum in Eq.~\eqref{eq:sm_ladder_hop} implements the Weyl ordering of the $k$-linear
SOC vertex:
\begin{equation}
  \frac12\{k_j,e^{i\theta}\}
  \;\longrightarrow\;
  \left(k_j+\frac{q_j}{2}\right)e^{i\theta}.
\end{equation}

For the numerical calculation the infinite ladder is truncated to $|n|\leq n_\mathrm{max}$
and convergence is checked by increasing both $n_\mathrm{max}$ and the momentum grid.
The resulting
finite matrix has dimension $4(2n_\mathrm{max}+1)$ at each reduced-zone momentum.
Within this
converged single-$q$ continuum model, the calculation retains the exchange winding and all SOC
winding matrix elements without projecting to the low-energy doublet and without pitch averaging.

The current vertices are obtained from the same ladder Hamiltonian by
\begin{equation}
  v_a(\mathbf k)=\frac{\partial H(\mathbf k)}{\partial k_a}, \qquad a=x,y .
\end{equation}
For example, the diagonal pieces contain
\begin{align}
  v_x^\mathrm{diag} &= \frac{\kappa_x}{m}\eta_0 + K^x\kappa_x\eta_x +
  \frac{K^z}{2}\kappa_y\eta_z + \lambda_A\eta_0\sigma_y + \lambda_B\eta_x\sigma_y, \\
  v_y^\mathrm{diag}
  &=
  \frac{\kappa_y}{m}\eta_0
  +
  K^x\kappa_y\eta_x
  +
  \frac{K^z}{2}\kappa_x\eta_z
  -
  \lambda_A\eta_0\sigma_x
  -
  \lambda_B\eta_x\sigma_x .
\end{align}
The off-diagonal $C$-channel velocity vertices follow by differentiating $S_C$:
\begin{align}
  \partial_{k_x}S_C(\boldsymbol\kappa,\mathbf a)
  &=
  a_z\sigma_x-a_x\sigma_z,
  \\
  \partial_{k_y}S_C(\boldsymbol\kappa,\mathbf a)
  &=
  a_z\sigma_y-a_y\sigma_z .
\end{align}

After diagonalizing the ladder Hamiltonian,
\begin{equation}
  H(\mathbf k)|u_\ell(\mathbf k)\rangle = E_\ell(\mathbf k)|u_\ell(\mathbf k)\rangle ,
\end{equation}
we evaluate the optical conductivity from the band-basis velocity matrix elements
\begin{equation}
  v_a^{\ell m}(\mathbf k) = \langle u_\ell(\mathbf k)|v_a(\mathbf k)|u_m(\mathbf k)\rangle .
\end{equation}
The interband kernel is constructed from all transitions $E_m(\mathbf k)>E_\ell(\mathbf k)$
with weight $f_\ell-f_m$, and is then broadened to obtain $\mathrm{Re}\,\sigma_{ab}(\omega)$.
Operationally,
\begin{equation}
  \mathrm{Re}\,\sigma_{ab}(\omega)
  \propto
  \int_\mathrm{RBZ}\frac{d^2k}{(2\pi)^2}
  \sum_{\ell\neq m}
  (f_\ell-f_m)
  \,v_a^{\ell m}v_b^{m\ell}\,
  \delta_\Gamma\!\left[\omega-(E_m-E_\ell)\right],
\end{equation}
with the proportionality convention identical to that used in the projected calculation.

This ladder calculation is therefore a direct check of the perturbative SOC analysis.
The
perturbative treatment separates the zero-Fourier SOC component from winding terms that mix
$\mathbf k$ with $\mathbf k\pm\mathbf q$; the superlattice calculation keeps these sidebands
explicitly and diagonalizes their hybridization nonperturbatively.
It therefore captures both
the form-preserving $q\to q_\mathrm{eff}$ shift at small SOC and the additional sideband/minigap
features that appear once the winding SOC terms are no longer negligible.

\begin{figure}[t]
  \centering
  \includegraphics[width=6.8in]{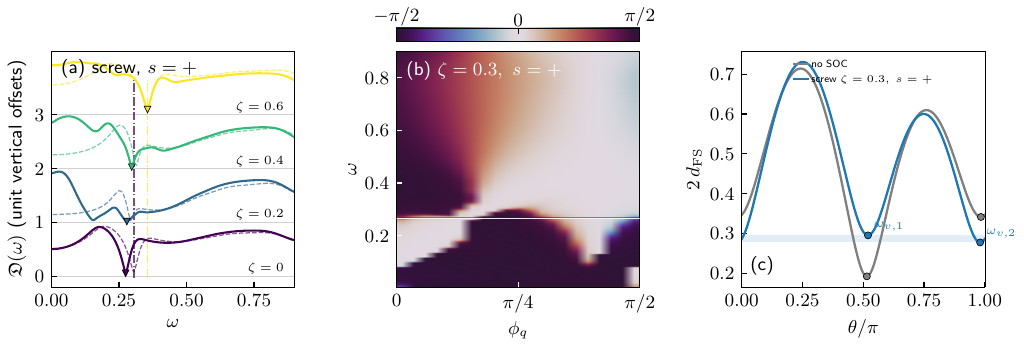}
  \caption{Robustness of the locked-to-tracking crossover against intrinsic SOC.
    Panels (a,b) use the four-band Bloch-ladder model ($J=20$, $\mu=2$, $C^x=0$,
    channel B); panel (c) uses the Fermi-circle approximation.
    (a)~Exact proper-screw dichroism at $\phi_q=\pi/6$ for $\zeta=0,0.2,0.4,0.6$
    (solid, unit-offset) and the split-$\bm q$ result (dashed).
    Markers are exact minima; dash-dot guides are the split-$\bm q$ predictions at
    $\zeta=0$ and $0.6$, whose drift follows the $\omega_c\propto q$ scaling of
    Eq.~\eqref{eq:omega_c_smallrho} under the substitution $\bm q\to\bm q_\mathrm{eff}$ of
    Eq.~\eqref{eq:soc_qeff}.
    (b)~Exact principal axis at $\zeta=0.3$, $s=+$; the horizontal guide marks
    $\omega_c$ at $\phi_q=\pi/6$; SOC mainly redistributes low-frequency weight between the
    orthogonal axes, while the crossover and high-frequency tracking are weakly affected.
    (c)~Fermi-circle gaps at $\zeta=0$ (grey) and $\zeta=0.3$, $s=+$ (blue);
  SOC reduces the distance between the two valleys.}
  \label{fig:soc_robustness}
\end{figure}

Fig.~\ref{fig:soc_robustness} summarizes the outcome.
For the proper screw shown, the
locked-to-tracking crossover survives as a single well-defined dichroism minimum as SOC increases
up to $\zeta=0.6$, the first-order split-$\bm q$ prediction departing from the exact result
only once $\zeta$ ceases to be small [Fig.~\ref{fig:soc_robustness}(a)].
At fixed $\zeta=0.3$
the exact angle-resolved response retains the same locked-to-tracking structure over a tetragonal
orientation sector [Fig.~\ref{fig:soc_robustness}(b)].
The largest SOC-induced changes occur at low frequency, where SOC redistributes absorption between
the two orthogonal principal axes and can interchange which one is dominant; this $\pi/2$ switch in
the reported axis does not remove the underlying helix-correlated principal-axis frame.
By contrast, the crossover scale and the high-frequency tracking regime are only weakly shifted.

To interpret the intermediate-frequency structure we return to the Fermi-circle direct gap
that sets the crossover scale (Sec.~\ref{app:omega_c_scaling}), compared with and without SOC
in Fig.~\ref{fig:soc_robustness}(c).
For a proper screw the first-order zero-Fourier SOC term
rotates the effective helix vector, $\bm q_\mathrm{eff}/q=\hat{\bm q}+s\zeta\,\hat{\bm q}_\perp$;
at the representative orientation $\phi_q=\pi/6$ this raises one dominant valley threshold
while lowering the other, so the two move closer together.
Because the corresponding valleys
favour competing polarization axes, their simultaneous optical contribution suppresses the net
dichroism and renders $\theta_+$ sensitive to small changes in frequency and helix orientation
across the intermediate-frequency window; the well-defined tracking relation
$\theta_+\simeq\pi/2-\phi_\mathrm{eff}$, with
$\phi_\mathrm{eff}=\phi_q+\arctan(s\zeta)$, re-emerges only at higher frequency.
The irregular
angular patches at low frequency in Fig.~\ref{fig:soc_robustness}(b) should not be read as
distinct phases: they occur where the interband absorption is weak and the principal axis is
correspondingly ill-defined.

\section{Parameters of the simulations}
\label{app:parameters}

All parameter values are given in the dimensionless units of the model and are chosen for
illustration rather than anchored to a specific material: they place the two Fermi-surface
gap scales of the problem in the same frequency window ($\rho=2K^xq/K^zk_F\simeq0.5$--$1.1$),
so that both the locked and the tracking regimes are visible in a single plot.
The scaling form of
Sec.~\ref{app:omega_c_scaling} translates the crossover frequency to any other parameter set.

\begin{table}[ht]
  \centering
  \caption{Parameters used in each figure.}
  \label{tab:params}
  \begin{tabular}{@{}llrrrrrrr@{}}
    \toprule
    Figure & Model & $m$ & $K^x$ & $K^z$ & $q$ & $\mu$ & $\Gamma$ & $T$ \\ \midrule
    Fig.~1 & $d$-wave & 1.0 & 1.0 & 0.3 & 0.4 & 3.0 & 0.08 & -- \\
    Fig.~1 & $g$-wave & 1.0 & 1.0 & 0.9 & 0.4 & 3.0 & 0.08 & -- \\
    Fig.~2 & $d$-wave & 1.0 & 1.0 & 0.3 & 0.2 & 3.0 & 0.02 & 0.05 \\
    Fig.~3 & $d$-wave & 1.0 & 1.0 & 0.3 & 0.2 & 3.0 & 0.02 & 0.05 \\
    Fig.~4 & $d$-wave & 1.0 & 1.0 & 0.3 & 0.2 & 2.0 & 0.02 & 0.05 \\
    Fig.~\ref{fig:conductivity} & $d$-wave & 1.0 & 1.0 & 0.3 & 0.2 & 3.0 & 0.05 & 0.05 \\
    Fig.~\ref{fig:conductivity} & $g$-wave & 1.0 & 1.0 & 0.9 & 0.2 & 3.0 & 0.05 & 0.05 \\
    Fig.~\ref{fig:afm_signatures} & $d$-wave & 1.0 & 1.0 & 0/0.3 & 0.04--0.40 & 3.0 & 0.05 & -- \\
    Fig.~\ref{fig:omega_c_mechanism} & $d$-wave & 1.0 & 1.0 & 0.3 & 0.2 & 3.0 & 0.08 & 0.05 \\
    Fig.~\ref{fig:omega_c_scaling} & $d$-wave & 1.0 & 1.0 & 0.3 & 0.04--0.46 & 3.0 & 0.08 & 0.05 \\
    Fig.~\ref{fig:multiband_robustness} & $d$-wave & 1.0 & 1.0 & 0.3 & 0.4 & 2.0 & 0.08 & 0.05 \\
    Fig.~\ref{fig:soc_robustness} & $d$-wave & 1.0 & 1.0 & 0.3 & 0.2 & 2.0 & 0.02 & 0.05 \\
    \bottomrule
  \end{tabular}
\end{table}
The exact four-band figures (Fig.~4, Fig.~\ref{fig:multiband_robustness} and
Fig.~\ref{fig:soc_robustness}) carry additional parameters beyond the projected-model columns
above; all use the exact four-band helix model at exchange $J=20$, deep in the projection
regime, and adopt $\mu=2$ (rather than the $\mu=3$ of the projected-model figures) so as to
sit at the quantitatively valid projection point $s=K^xk_F^2/2E_0=0.1$.
For the multiband check
(Fig.~\ref{fig:multiband_robustness}) the sublattice hybridization is $C^x=0$ in panels (a) and (c)
and $C^x=J/2$ in panel (b), with $\mu=2$ the working point of panels (b) and (c) while panel~(a)
sweeps $\mu$ to vary $s$.
The intrinsic-SOC results of Fig.~4 and Fig.~\ref{fig:soc_robustness}
set $C^x=0$ and use the dimensionless SOC strength $\zeta=2\bar\alpha/(K^xq)$:
Fig.~\ref{fig:soc_robustness}(a) shows the proper-screw dichroism waterfall at $\zeta=0,0.2,0.4,0.6$
and $\phi_q=\pi/6$; panel (b) fixes $\zeta=0.3$ at
$0\le\phi_q\le\pi/2$, and panel (c) compares the $\zeta=0$ and $\zeta=0.3$ Fermi-circle gaps at
$\phi_q=\pi/6$.
The chemical potential is calibrated so that the Fermi momentum of the low-energy manifold
matches the projected value $k_F=\sqrt{2m\mu}$, while the first-order curves use the projected
split-$\bm q$ model at the same $k_F$.

\bibliography{am_textures}